\documentclass{elsart}

\usepackage{graphicx}

\usepackage{amssymb}
\DeclareRobustCommand*\cal{\relax\mathcal}

\begin{document}

\begin{frontmatter}



\title{Combinatorial and approximative analyses\\
in a spatially random division process}


\author[JAIST]{Yukio Hayashi}, 
\author[JAIST]{Takayuki Komaki}
\author[Kanagawa]{Yusuke Ide}
\author[Meiji]{Takuya Machida}
\author[Yokohama]{Norio Konno}

\address[JAIST]{Japan Advanced Institute of Science and Technology, 
Ishikawa 923-1292, Japan}
\address[Kanagawa]{Kanagawa University,
Kanagawa, 221-8686, Japan}
\address[Meiji]{Meiji University,
Kanagawa, 214-8571, Japan}
\address[Yokohama]{Yokohama National University,
Kanagawa, 240-8501, Japan}

\begin{abstract}
For a spatial characteristic, 
there exist commonly fat-tail frequency distributions of
fragment-size  and -mass of glass,
areas enclosed by city roads, and
pore size/volume  in random packings.
In order to give a new analytical approach for the distributions, 
we consider a simple model 
which constructs a fractal-like hierarchical network 
based on random divisions of rectangles.
The stochastic process makes a Markov chain
and corresponds to directional random walks with
splitting into four particles.
We derive a combinatorial analytical form and its continuous
approximation for the distribution of rectangle areas, 
and numerically show a good fitting with the actual distribution
in the averaging behavior of the divisions.
\end{abstract}

\end{frontmatter}

\newpage
\section{Introduction}
Historically, 
a discussion for the origin of skew distribution that appears 
in sociological, biological, and economic phenomena 
goes back to the Simon's stochastic model \cite{Simon55}.
At the beginning of this century, 
the evidences of fat-tail degree distribution have been also
observed in many real networks \cite{Newman06}
through computer analyses for large data.
Some cases look like a power-law distribution
while some other cases a lognormal distribution,
it is difficult to discriminate these distributions in general.
Since the tail in a lognormal distribution resemble 
power-law behavior, only the part may be observed.
Moreover, 
the generation mechanisms of the   distributions are not
exclusive but intrinsically connected in preferential attachment, 
multiplicative process, and other models \cite{Mitzenmacher04a}.
On a stochastic process of geometric Brownian motion, 
a model of double Pareto distribution generates 
a lognormal body and Pareto tail in the continuous
distribution \cite{Mitzenmacher04b}\cite{Reed04}.
On other process of iterative division of cells, 
the frequencies experimentally follow 
such distributions of 
fragment-size and -mass of glass \cite{Katsuragi04}\cite{Ishii92},
areas enclosed by city roads
\cite{Lammer06}\cite{Masucci09}\cite{Chan11},
pore size/volume in
random packings \cite{Delaney08}\cite{Dodds03},
and areas enclosed by edges in models of
urban street patterns \cite{Barthelemy08} and
geographical networks \cite{Lee12}.
In addition, 
the distribution in fragmentation of glass
changes from a lognormal to a power-law-like
according to low and high impacts \cite{Katsuragi04},
which determine the limitation of breakable sizes.
It is worthwhile to study a mechanism in abstract models
for generating the similar distributions 
in spite of different
physical quantities and operations in a variety of research fields:
socio-economic, material, computer, and physical sciences, 
even if we apart from the reproduce of realistic phenomena 
and do not insist on the detail process or 
the macroscopic properties exactly in broken fragmentation of glass.

For crack patterns, 
a random tessellation model has been proposed \cite{Nagel05}.
It is based on a stochastic point process, consisting of the 
division of a randomly chosen face (cell) 
according to its life-time
by adding a random line segment.
The distribution of tessellations is invariant for an appropriate
rescaling, whose characteristic is called 
stable with respect to iteration (STIT).
In addition, the length distribution of segments is analytically 
obtained in a classification of several types of segments 
\cite{Nagel09}\cite{Thale09}.
However, the distribution of areas enclosed by segments is not 
derived.
The adjustment of life-time for each cell is also not easily 
applicable even in the sophisticated mathematical model of STIT, 
when we consider a construction of network 
in a procedural manner such as load balancing in a territory of node.

Thus, one of the issues is a theoretical analysis
for the distribution of areas in a fractal-like structure.
The presence of hierarchy and scaling law 
is important \cite{Kalapala06}
for understanding the universal mechanism to generate
such a structure.
Through a mathematical model, 
we focus on the 
spatial phenomena based on dividing by four with a randomness
for the simplicity with analogous structures to road networks.
However, for the iterative division processes, 
only a few mathematical models are known.
In random tessellations, 
although the cell-selection and cell-division rules are classified 
into equally-likely, area-weighted, perimeter-weighted, 
corner-weighted, and so on,
the length distributions in one-dimension are merely analyzed 
for some simple rules \cite{Cowan10}.
In a quadtree model characterized as a typical road network, 
the shortest paths and maximum flow are analyzed
\cite{Eisenstat11}, but the distribution of areas 
is not discussed.

On the other hand, we recently proposed
{\em multi-scale quartered} (MSQ) networks based on 
a self-similar tiling by equilateral triangles
or squares \cite{Hayashi09,Hayashi10}.
This model is constructed by iterative division of faces, 
and is also suitable 
for the analysis of depth distribution of layered areas
in a framework of Markov chain \cite{Hayashi11}.
Moreover, from an application point of view, 
the MSQ networks without hub nodes
have several advantages of the strong robustness of
connectivity against node removals by random failures and intentional
attacks, the bounded short path as $t=2$-spanner
\cite{Karavelas01}, 
the efficient face routing by using only local information, and 
a scalable load balancing performed
by the divisions of the node's territory for increasing communication
or transportation requests.
However, due to the self-similar tiling 
in the scalably growing MSQ networks, 
the position of a new node is restricted
on the half-point of an edge of the chosen face, and the
link length is proportional to $(\frac{1}{2})^{H}$
where $H$ is the hierarchical depth number of divisions.
The restriction is unnatural 
for many division processes in physical or social 
phenomena.
Thus, we generalize the divisions of squares
to ones of rectangles with any link lengths
instead of the iterative halvings.

The organization of this paper is as follows.
In Sec. \ref{sec2}, we introduce a generalization
of MSQ network model.
In Sec. \ref{sec3}, for the distribution of areas,
we derive the exact solution on a combinatorial analysis.
We point out that the behavior of divisions 
is equivalent to directional random walks 
with splitting into four particles.
The  representation of random walks with splitting
gives us an inspiration for the combinatorial analysis.
However, this approach is limited for application 
to very small networks.
In Sec. \ref{sec4}, 
we consider a continuous approximation of the 
distribution of areas for large networks.
We decompose the distribution function
into two components of Poisson and gamma distributions, 
and numerically investigate the fitting of the mixture distribution
by the two components
with the actual distribution of areas in the divisions
for generating a fractal-like network structure.
In Sec. \ref{sec5}, we summarize these results.

\newpage
\section{Division process} \label{sec2}
We consider a two-dimensional $L \times L$ square, 
whose lattice points $(0,0)$, $(0,1)$, $\ldots$, $(0,L)$, 
$\ldots$, $(L,0)$, $(L,1)$, $\ldots$, $(L,L)$ 
in the $x$-$y$ coordinates 
give the feasible setting positions of nodes. 
Initially, there exist only the outer square with 
four corner nodes and edges of the length $L$. 
The following model generalizes the MSQ network 
\cite{Hayashi09,Hayashi10,Hayashi11} 
from recursive divisions of squares
to ones of rectangles.
In the MSQ network based on a self-similar tiling by squares, 
the division is restricted at the half point of an edge, 
therefore we extend the division to that at a cross point 
of the vertical and horizontal segments on the square lattice. 
The case of $L \rightarrow \infty$ 
gives a general position for the division point. 
Varying the value of $L$ controls 
the limitation of divisible size relatively, 
as similar to low and high impacts in the fragmentation 
of glass \cite{Katsuragi04}.

\begin{figure}[htb]
  \begin{center}
    \includegraphics[height=50mm]{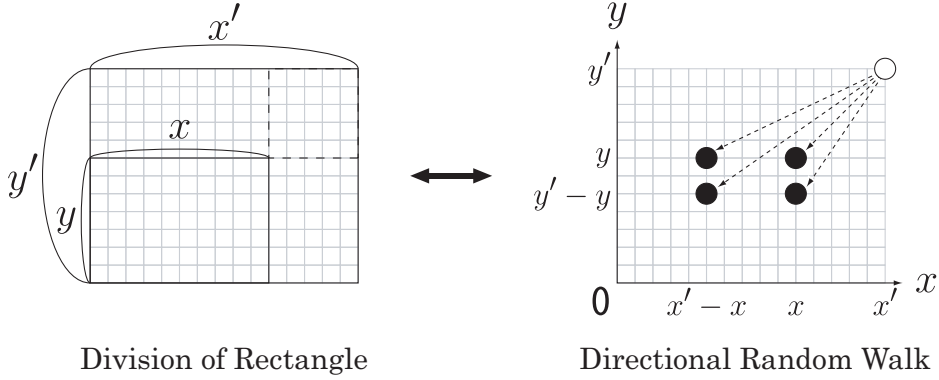}
  \end{center}
\caption{Correspondence between the division of a rectangle 
and the directional random walks with splitting.
Vertical and horizontal thin lines at even intervals
are the segments of an $L \times L$ lattice.}
\label{fig_mapping}
\end{figure}

The proposed network is iteratively constructed for a given $L$ 
as follows. 
At each time step, 
a rectangle is chosen uniformly at random (u.a.r), 
and it is divided into four smaller rectangles.
Then, the smaller rectangle with an area $x \times y$ 
($x,y$ denote the two edge lengths) is generated from 
the chosen rectangle with an area $x' \times y'$. 
Simultaneously, rectangles with the areas $(x'-x) \times y$,
$x \times (y'-y)$, and $(x'-x) \times (y'-y)$ are generated.
Here, 
two division axes are chosen u.a.r from 
the horizontal and vertical segments of an $L \times L$ lattice 
(see the left of Fig. \ref{fig_mapping}).
In other words, each edge length 
$x,y \in Z_{+} = \{ 1, 2, \ldots \}$ 
is randomly chosen as a positive integer in 
$x+1 \leq x' \leq L$ and $y+1 \leq y' \leq L$.
The stochastic network generation makes a Markov chain.
The state is represented by a vector 
$(n_{11}, \ldots, n_{xy}, \ldots, n_{LL})$, 
where $n_{xy}$ denotes the number of rectangle 
with the area $x \times y$ (The Markov chain is degenerately simplified 
by ignoring the difference of areas in subsection 4.1 
for discussing the distribution of layers 
defined by the depth of divisions.).
The stochastic process is characterized by the fact that 
the transition probability to divide a rectangle with the area 
$x \times y$ is not fixed but proportional to $n_{xy}$ 
because of the uniformly at random selection of a face. 
In other words, the probability
depends on a sequence of chosen rectangles during the transition 
until 
a final absorbing state for the indivisible width 
$x = 1$ or $y = 1$. 
We remark that the minimum edge length $x = 1$ or $y = 1$ 
bounds the number of the states finitely, 
while the MSQ networks \cite{Hayashi11} have 
the infinite states without a limitation of the subdivision.
The scaling relation of the maximum iteration time is 
numerically obtained as 
$T_{max} \sim L^{1.91}$ in the averaging of absorbing states.
Although this paper discusses a simple case of 
uniformly random selections of rectangles and of division points 
in order to deduce the distribution of areas, 
the selections can be extended to other cases,
e.g., according to a given population in a territory of node
for real statistical data.
Note that the above process is different from the Galton-Watson type
branching process with a time-independent probability for generating
offspring \cite{Liggett99}.

\section{Combinatorial analysis} \label{sec3}
As shown in Fig. \ref{fig_mapping}, 
the recursive generation process can be regarded as 
directional random walks of   particle 
with splitting into four copies 
in the framework of two-dimensional cellular automaton (CA), 
when a pair $(x, y)$ of edge lengths of a rectangle 
is corresponded to the position of particle 
in the $x$-$y$ coordinates. 
A particle is randomly chosen at a time step, 
and moves toward smaller coordinate values
from $(x', y')$ to $(x, y)$, where $x < x'$ and $y < y'$, 
until reaching the boundary at $x = 1$ or $y = 1$.
We emphasize that 
this type of CA with splitting 
differs from the asymmetric simple exclusion process 
(ASEP) \cite{Derrida93} 
and the contact (or voter) model \cite{Liggett99}, 
since there are no spatial exclusions 
and no interaction between any particles.

This representation of random walks with splitting is 
inspirational for deriving a combinatorial analytic form 
of the distribution of areas.
We consider the number $n_{xy}$ of particles at $(x,y)$, 
equivalently the number of 
rectangles with the area $x \times y$.
Remember that $x$ and $y$ are integers.
The average behavior is described by the following 
system of difference equations for $2 \leq x,y \leq L-1$
with the sum by taking over the integers 
$x+1 \leq x' \leq L$ and $y+1 \leq y' \leq L$,
\begin{equation}
  \Delta n_{xy} = -p_{xy} + \sum_{x',y'} 
	\frac{4 p_{x'y'}}{(x' -1)(y' -1)}, \label{eq1}
\end{equation} 
where 
$\Delta n_{xy}$ is the average difference of $n_{xy}$ in one step, and 
$p_{xy} \stackrel{\rm def}{=} 
n_{xy} / \sum_{x" > 1 y" > 1} n_{x"y"}$ is the existing probability
of a particle at $(x,y)$.
The factor $4$ in the numerator of 
right-hand side of Eq.(\ref{eq1})
is due to feasible positions of the $x \times y$
at left/right and upper/lower corners in the division
of $x' \times y'$.
The denominator is the combination number for the 
relative positions of emanating particles 
in the intervals $[1, x'-1]$ and $[1, y'-1]$,
and is equivalent to the number for selecting two axes in the 
division of rectangle with the area $x' \times y'$.

From $\Delta n_{xy} = 0$ in Eq.(\ref{eq1}), we derive
\[
  p_{L-1L-1} = \frac{4 p_{LL}}{(L-1)^{2}},
\]
\[
  p_{xL-1} = p_{L-1y} = 
	\frac{4 p_{LL}}{(L-1)^{2}}, \;\;
	x > 1, y > 1, 	
\]
\[
  p_{L-2L-2} = \left( 1 + \frac{4}{(L-2)^{2}} \right)
	\frac{4 p_{LL}}{(L-1)^{2}}.
\]
In general, we obtain the solution 
by applying the above in decreasing order of $x$ and $y$
recursively.
\begin{equation}
  p_{xy} = \left\{ 1 + \sum_{{\cal P}}
	\left( \Pi_{i=1}^{l} 
	\frac{4}{(x_{i}-1)(y_{i} -1)} 
	\right) \right\}
	\frac{4 p_{LL}}{(L-1)^{2}}, 
\label{eq_pxy}
\end{equation}©
where $\sum_{{\cal P}}$ denotes the sum for a set of paths 
through the points
$(x_{1},y_{1}),(x_{2},y_{2}), \ldots, (x_{l},y_{l})$,
with $x_{i}, y_{i} \in Z_{+}$, 
$x < x_{1} < x_{2} < \ldots < x_{i} < \ldots x_{l} \leq L-1$, and  
$y < y_{1} < y_{2} < \ldots < y_{i} < \ldots y_{l} \leq L-1$ 
in all combinations of 
$l = 1, 2, \ldots \min\{ L-1-x, L-1-y \}$.

By substituting the solution $p_{xy}$ of Eq.(\ref{eq_pxy}) 
into the following right-hand sides, 
\[
\begin{array}{lll}
  n_{1y} & = & \sum_{x' > 1,y' > y} 
	\frac{4 p_{x'y'}}{(x'-1)(y'-1)}, \\
  n_{x1} & = & \sum_{x' > x,y' > 1} 
	\frac{4 p_{x'y'}}{(x'-1)(y'-1)}, \\
  n_{11} & = & \sum_{x' > 1,y' > 1} 
	\frac{4 p_{x'y'}}{(x'-1)(y'-1)},
\end{array}
\]
we obtain the distribution $P(A)$ of rectangles with the area $A$.
The sum is taken over the positive integers $x',y' \leq L$, 
$n_{11}$, $n_{x1}$, and $n_{1y}$ denote the numbers of the 
finally remaining rectangles with the areas $1 \times 1$, $x \times 1$, 
and $1 \times y$, which are no more divisible.
Note that the unknown factor $p_{LL}$ disappears 
by the numerator and the denominator in all of 
$P(1) = n_{11} / {\cal N}$ and        
$P(x) = (n_{x1} + n_{1x}) / {\cal N}$,
where 
${\cal N} = n_{11} + \sum_{x''  > 1} n_{x'' 1} 
+ \sum_{y''  > 1} n_{1y'' }$
denotes the total number of the divided rectangle faces.

\begin{figure}[htb]
  \begin{center}
	\includegraphics[height=115mm,angle=-90]{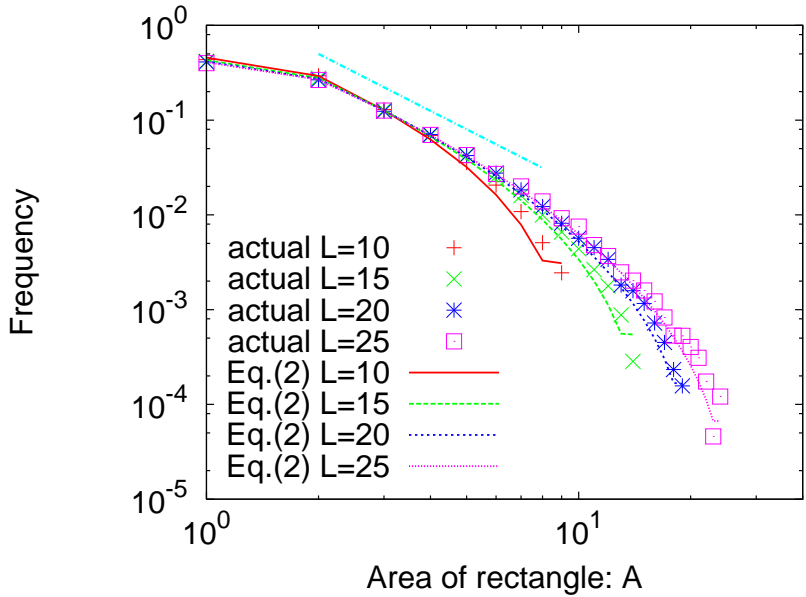}
  \end{center}
\caption{(Color online) Distribution of areas: 
$1 \times 1, 1 \times 2, \ldots, 1 \times L-1$ 
in the extreme rectangles
for $L = 10, 15, 20,$ and $25$ from left to right.
The (plus, cross, star, and rectangle)
marks show the averaged result 
in 100 samples of the actual divisions, 
and the corresponding (red, green, blue, and magenta)
lines show the solution of Eq.(\ref{eq_pxy}) on 
the combinatorial analysis.
The short (cyan) segment guides the slope of $-2$
corresponded to the exponent in broken fragments 
of glass \cite{Katsuragi04} 
and city roads \cite{Lammer06,Masucci09}.
} \label{fig_limit_dist}
\end{figure}

Figure \ref{fig_limit_dist} shows the distribution of areas 
with width one.
Our solution denoted by lines is almost completely fitting 
with the actual distribution denoted by marks.
The part of linear tail becomes longer, as $L$ is larger.
Note that these distributions in Fig. \ref{fig_limit_dist}
are estimated well by lognormal functions in this entire 
range of $A$.

\section{Continuous approximation} \label{sec4}
Since it possibly causes a combinatorial explosion 
to calculate the extreme distribution of areas: 
$1 \times 1, 1 \times 2, \ldots, 1 \times L-1$ 
at the absorbing states in the 
Markov chain, 
the application of Eq.(\ref{eq_pxy})
is restricted for a very small $L$. 
In order to analyze the distribution of areas for a large $L$, 
we approximate the process 
to be divisible at any positions on two edges of a rectangle, 
by ignoring the restriction to the 
segments on an $L \times L$ square lattice.
For $l = 1,2,\ldots$ until the maximum layer at a given time, 
we consider 
the sum of the product of $p_{l}$ and $g_{2l}(\log A)$, 
which denote the frequencies of layer $l$ 
and of area $A$ in the layer $l$.
Here, the number $l$ represents the depth of divisions.
We derive these frequencies separately.
Numerical simulations in subsection 4.3 
show a good fitting of the mixture distribution 
$\sum_{l} p_{l} g_{2l}(\log A)$ with the actual distribution of areas 
in the average behavior of the divisions.

\subsection{Distribution of layers}
\begin{figure}[htb]
  \begin{center}
    \includegraphics[height=73mm]{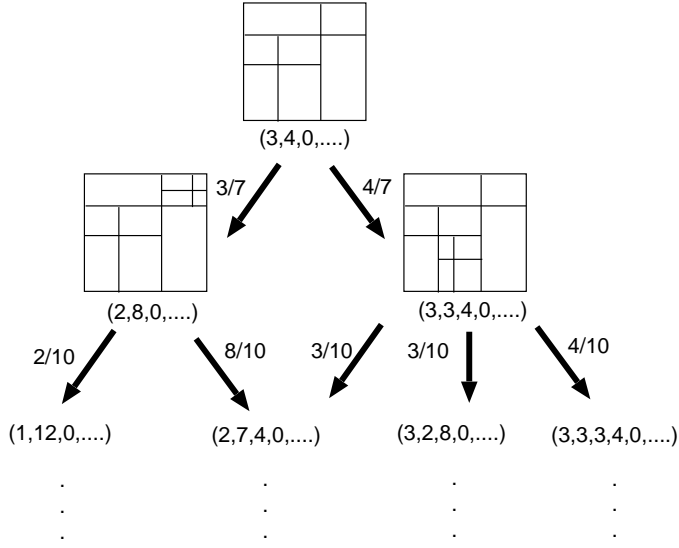}
  \end{center}
\caption{Branching tree diagram of the state vector 
$(n_{1}, n_{2}, \ldots)$ for the division process at 
$t = 2,3,4$ steps (from top to bottom).
Here, $n_{l}$ is the number of faces on the $l$-th depth,
when the difference of areas is ignored in the count.
Each fraction denotes the transition probability.
Note that two transitions until $t=2$ through 
$(4,0,\ldots)$ at $t=1$ and the initial square
are trivial.} \label{fig_branch}
\end{figure}

In this subsection, we ignore the difference of areas in each layer, 
and treat only the number of faces.
Then, the stochastic division process characterized as a Markov chain 
is simplified.
Figure \ref{fig_branch} shows the state transitions 
in the first few steps.
We consider the number $n_{l}(t)$ of faces 
in the $l$-th layer at a time step $t$,
and derive approximative solutions for 
the existing frequency of faces in the layer.

As shown in \cite{Hayashi11}, the averaging behavior of difference 
\begin{equation}
  \Delta n_{l} \stackrel{\rm def}{=} n_{l}(t+1) - n_{l}(t), 
   \label{eq_delta_nl1}
\end{equation}
can be written to
\begin{equation}
  \Delta n_{l} = m p_{l-1}(t) - p_{l}(t), \label{eq_delta_nl2}
\end{equation}
since a face in the layer $l$ chosen with the probability $p_{l}(t)$
is divided into $m=4$ smaller ones which belong to the layer $l+1$,
therefore a face in the layer $l-1$ contributes
to increase the number of faces in the layer $l$.
Note that the simultaneously created $m$ faces at a time 
belong to a same layer even with different areas.
For a large $t$, by noticing $n_{l}(t) = {\cal N}(t) p_{l}(t)$ and
substituting 
${\cal N}(t) = \sum_{l} n_{l} = 1 + (m-1) t \approx (m-1) t$
into the right-hand side of Eq. (\ref{eq_delta_nl1}), it is
\[
  \begin{array}{lll}
  \Delta n_{l} & = & (m-1)(t+1) p_{l}(t+1) - (m-1)t p_{l}(t), \\
               & = & (m-1)t [ p_{l}(t+1) - p_{l}(t) ] + (m-1) p_{l}(t+1).
  \end{array}
\]
Using $p_{l}(t+1) \approx p_{l}(t)$
because of $t+1 \approx t \gg 1$,
Eq. (\ref{eq_delta_nl2}) is rewritten to

\begin{equation}
  p_{l}(t+1) - p_{l}(t) = - \frac{m}{(m-1) t} 
   \{ p_{l}(t) - p_{l-1}(t) \}, \label{eq_difference}
\end{equation}
where $p_{0} \equiv 0$ is assumed for convenience.
The solution of Eq.(\ref{eq_difference}) is not easily 
derived even from a formal representation 
by using a generating function 
because of the combinatorial explosion
involved with very complicated recursive operations
\cite{Hayashi11}.
For the $m=2$ divisions, the difference equation 
(\ref{eq_difference}) is equivalent to Eq.(14) in \cite{Cowan10}
for a crack    model, 
however it differs to consider 
a rescaling length factor at each time step
and to analyze a cumulative distribution 
in the one-dimensional case.

On the other hand, 
we also derive the following expression \cite{Hayashi11} 
by using a model in an interacting infinite particle system, 
\begin{eqnarray}
  \frac{d n_{l}}{d \tau} & = & 
	m n_{l-1} - n_{l}, \;\;\; l \geq 2, \label{eq_diff_tau}\\
  \frac{d n_{1}}{d \tau} & = & - n_{1}.
\end{eqnarray}
The solution is 
\begin{equation}
  n_{l} = \frac{(m \tau)^{l-1}}{(l-1)!} e^{-\tau}, \label{eq_sol_nl}
\end{equation}
\begin{equation}
  {\cal N}(\tau) = e^{(m-1)\tau}, \label{eq_sol_N}
\end{equation}
\begin{equation}
  p_{l} = \frac{(m \tau)^{l-1}}{(l-1)!} e^{-m \tau},
   \label{eq_sol_pl}
\end{equation}
where $\tau \geq 0$ has a logarithmic timescale 
from the relation $1 + (m-1) t = e^{(m-1)\tau}$
of the total number of faces.
Note that the distribution of Eq.(\ref{eq_sol_pl}) coincides 
with the solution of Eq.(\ref{eq_difference}) asymptotically 
after a huge time \cite{Hayashi11}.
This form of $n_{l}$ in Eq.(\ref{eq_sol_nl})
can be applied for calculating 
a fractal dimension at a proper time 
in the MSQ network based on a self-similar tiling, 
and extended to a preference model for selecting a face 
in Appendixes A and B.

\subsection{Distribution of areas in the $l$-th layer} 
Remember that, 
at each time step, a rectangle is chosen uniformly at random 
(u.a.r).
For the division, vertical and horizontal axes are also 
chosen u.a.r
from the segments on an $L \times L$ square lattice. 
We focus on a set of rectangle faces on only the $l$-th layer 
generated in some steps.
After $l$-time selections, 
the subdivided face belongs to the layer $l$.
The area $S_{l}$ is given by the product
of shrinking rates $0 < X_{i}, Y_{i} < 1$, $i = 1, 2, \ldots, l$, 
for two edges of rectangle, 
\[
  S_{l} = \Pi_{i= 1}^{l} X_{i} Y_{i} L^{2},
\]
where $X_{i}$ and $Y_{i}$ are rational numbers, 
and $S_{l}$ is a positive integer in the division process, 
strictly speaking.

As an approximation for a large $L$, 
we assume 
that the random variables $X_{i}$ and $Y_{i}$ follow 
a $(0, 1)$ uniform distribution.
Then we define a variable 
$x \stackrel{\rm def}{=} - \log (S_{l} / L^{2}) =
	- \sum_{i} ( \log X_{i} + \log Y_{i} )$
in the range of equivalent relation 
$x \geq 0 \Leftrightarrow L^{2} \geq S_{l}$, 
the probability of $x$ follows a gamma distribution 
\begin{equation}
  g_{2l}(x) = e^{-x} \frac{x^{2l-1}}{(2l -1 )!}, 
\label{eq_gamma_dist}
\end{equation}
Here, the mean $\mu$ and the variance $\sigma^{2}$ of $\log X_{i}$
are 
\[
  \mu = \int_{0}^{1} \log X dX = -1,
\]
\[
  \sigma^{2} = \int_{0}^{1} (\log X - \mu)^{2} dX = 1.
\]
Therefore, by a central-limit theorem, 
we have a normal distribution $N(0,1)$ asymptotically 
\[
  \frac{\log S_{l} -(\log L^{2} - 2l)}{\sigma \sqrt{l}} 
	\rightarrow N(0,1),
\]
\[
  \log L^{2} - 2l
	= \log \left( \frac{L}{e^{l}} \right)^{2}.
\]
In other words, the average shrinking rate of edge is $1/e$ 
for each division of a rectangle, 
and it is slightly smaller than $1/2$ 
for that of a square in the MSQ network.
We can easily transform $g_{2l}(x)$ to the function of 
$\log S_{l}$ by using the shift of 
$x = 2 \log L - \log S_{l}$ for a constant $L$.

\subsection{Simulation for the mixture distribution} 
In numerical simulation, 
we investigate the distributions decomposed 
into the approximative $p_{l}$ and $g_{2l}$, 
and discuss the condition for a good fitting 
to each of components in 
the actual distributions for the divisions of rectangles.
Indeed, our approximation of $\sum_{l} p_{l} g_{2l}$ shows 
reasonable agreement with the actual distribution of areas.
Here, we consider the complementary cumulative distribution 
(CCD) of areas with a merit of smoothing effect, 
because the frequency distribution itself has a huge variety of areas 
especially for a large $L$, 
therefore it is practically impossible to gather these samples
in a proper frequency.

\begin{figure}[htb]
 \begin{minipage}[htb]{.47\textwidth}
   \includegraphics[height=55mm]{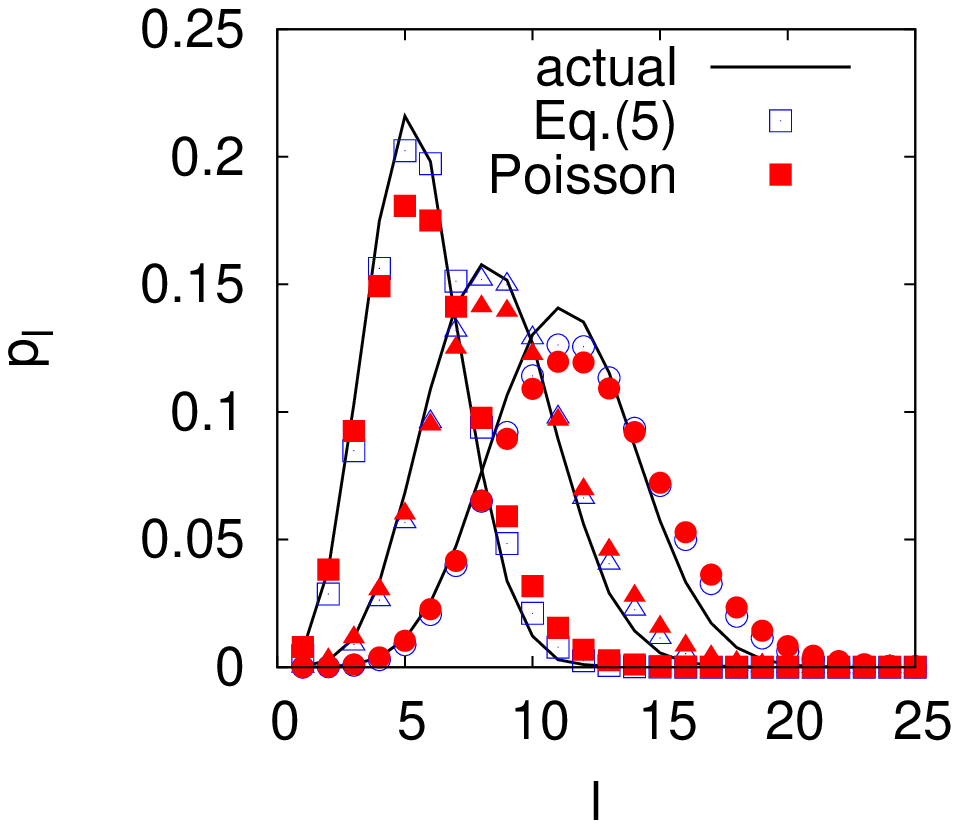}
     \begin{center} (a) $L = 10^{8}$ \end{center}
 \end{minipage} 
 \hfill 
 \begin{minipage}[htb]{.47\textwidth}
   \includegraphics[height=55mm]{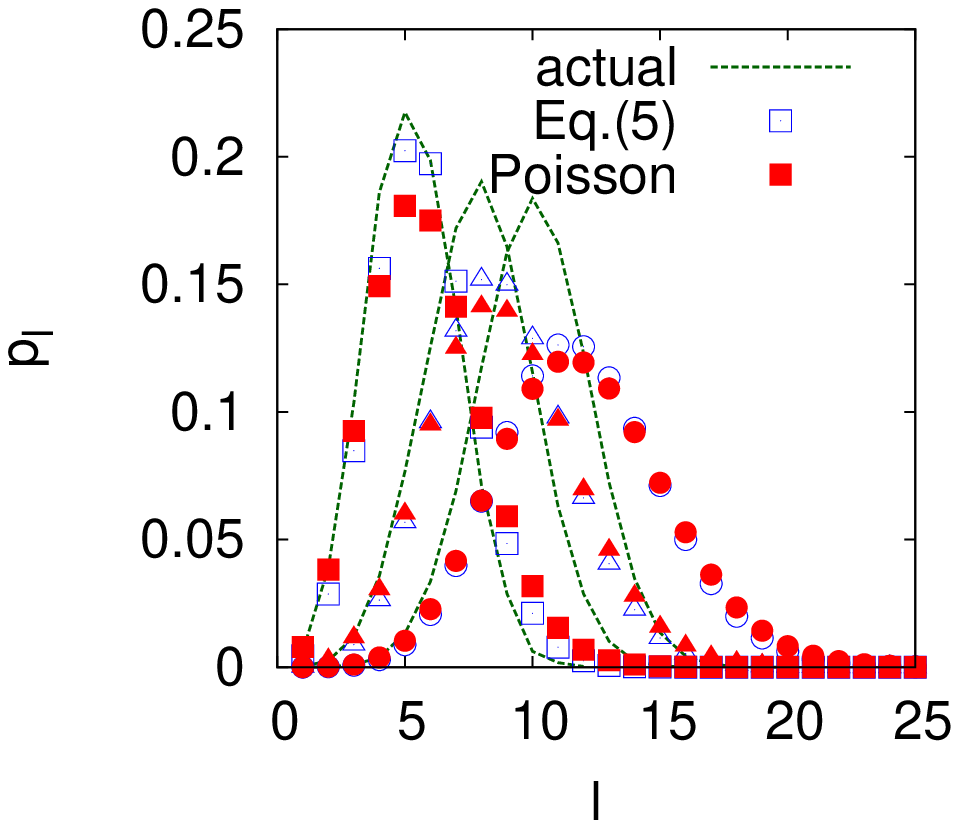}
     \begin{center} (b) $L = 10^{5}$ \end{center}
 \end{minipage} 
 \begin{minipage}[htb]{.47\textwidth}
   \includegraphics[height=55mm]{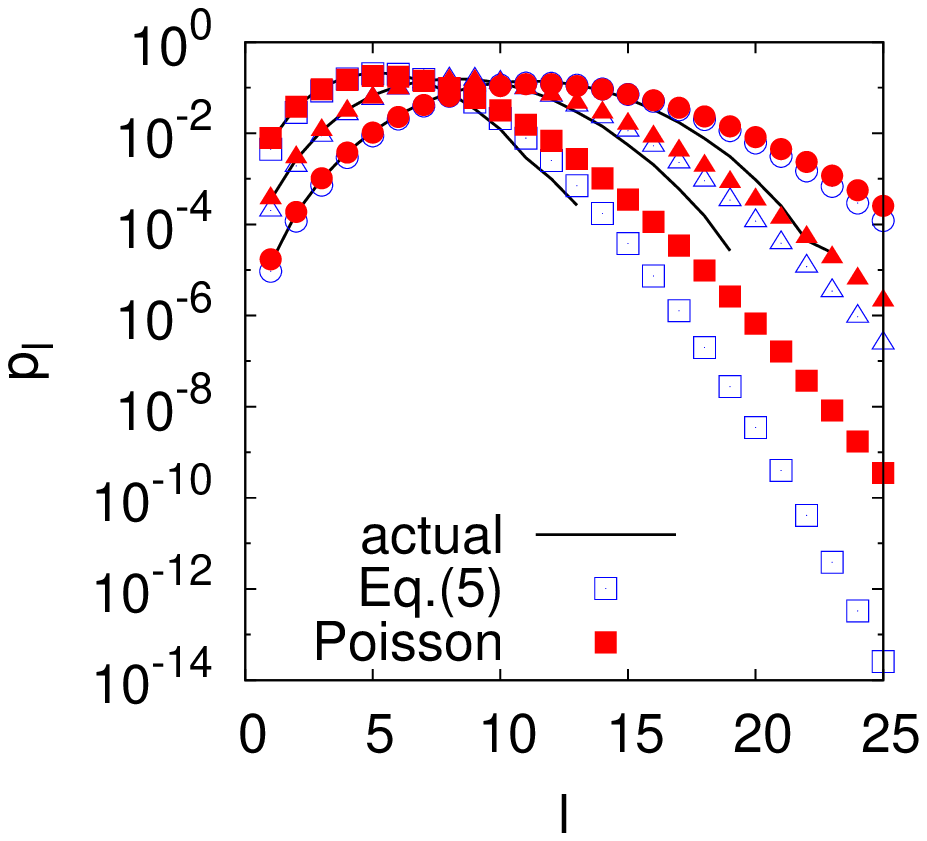}
     \begin{center} (c) $L = 10^{8}$ \end{center}
 \end{minipage} 
 \hfill 
 \begin{minipage}[htb]{.47\textwidth}
   \includegraphics[height=55mm]{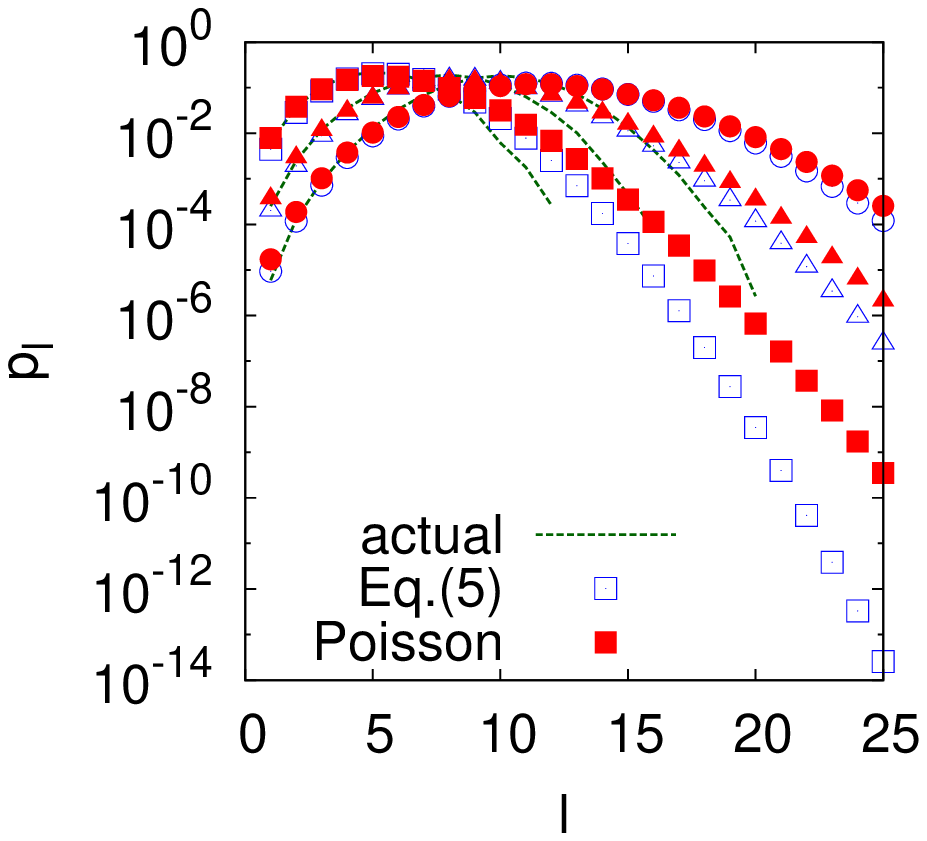}
     \begin{center} (d) $L = 10^{5}$ \end{center}
 \end{minipage} 
\caption{(Color online) Distributions $p_{l}$ of layers at time steps 
$t = 50, 500,$ and $5000$ from left to right.
The (black) solid and (green) dashed lines correspond to 
the averaged result by 100 samples of 
the actual divisions of rectangles for (a) $L = 10^{8}$
and (b) $L = 10^{5}$, respectively.
The semi-log plots are shown in (c) and (d) for investigating 
the fitting in the part of tail.
The open and closed marks correspond to 
the solution of Eq.(\ref{eq_difference})
and Poisson distribution in Eq.(\ref{eq_sol_pl}).
The discrepancy of positions between the open marks 
and the lines is due to the effect of width one 
in the extreme rectangles.
Note that these marks for $L = 10^{8}$ and $10^{5}$ 
are the same at each time step, and that only the 
solid and dashed lines are different in (a) and (b)
or (c) and (d).
} \label{fig_compare_pl}
\end{figure}

Figure \ref{fig_compare_pl} shows 
the distribution $p_{l}$ 
of layers at time steps $t= 50, 500,$ 
and $5000 \ll T_{max} \sim L^{1.91}$ 
for $L = 10^{8}$ and $10^{5}$.
By the effect of width one, 
there exists a gap between 
the solution of difference equation(\ref{eq_difference}) 
denoted by open marks 
and the actual distribution denoted by lines,
although these distributions almost coincide 
in the MSQ networks based on a self-similar tiling
\cite{Hayashi11}.
The gap becomes slightly larger as $L$ is smaller 
in Fig. \ref{fig_compare_pl}(b)
and $t$ is larger further to the right, 
because the effect tends to appear in more 
coarse-grained divisions and a deeper layer.
With the growing of the total number 
${\cal N} = 1 +(m-1)t$ of faces, 
the depth of a face tends to be deeper 
as the time step $t$ is larger.
In addition, we note the expectation 
$\langle l \rangle \sim \log t$ \cite{Hayashi11}.
The Poisson distribution of Eq.(\ref{eq_sol_pl}) 
denoted by closed marks has a slightly larger gap than 
the solution of difference equation(\ref{eq_difference}) 
denoted by open marks.
In the semi-log plots of Fig. \ref{fig_compare_pl}(c)(d),  
we also investigate the fitting of the tails.
In the tails, 
the discrepancy between the actual distribution 
and our approximation appears for the large $L = 10^{8}$
as $t$ is larger further to the right, 
while these distributions do not fit any longer for the 
small $L = 10^{5}$.
Here, in order to keep    the accuracy of approximation, 
we set the initial condition of $\{ p_{l}(0) \}$
by the existing frequency of faces in each layer 
at $t=5$ that is explicitly determined from the branching tree diagram
with the transition probability as shown in Fig. \ref{fig_branch}.
Note that the calculation becomes more complex as $t$ is larger, 
although a higher accuracy is expected.

\begin{figure}[htb]
 \begin{minipage}[htb]{.47\textwidth}
   \includegraphics[height=55mm]{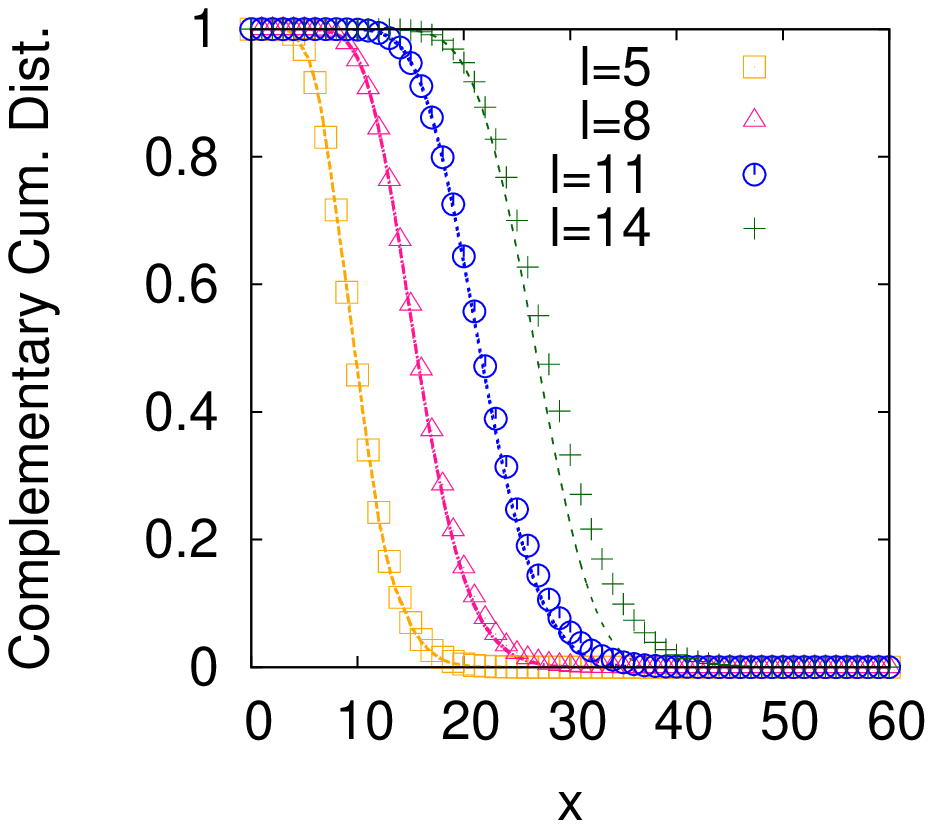}
     \begin{center} (a) $L = 10^{8}$ \end{center}
 \end{minipage} 
 \hfill 
 \begin{minipage}[htb]{.47\textwidth}
   \includegraphics[height=55mm]{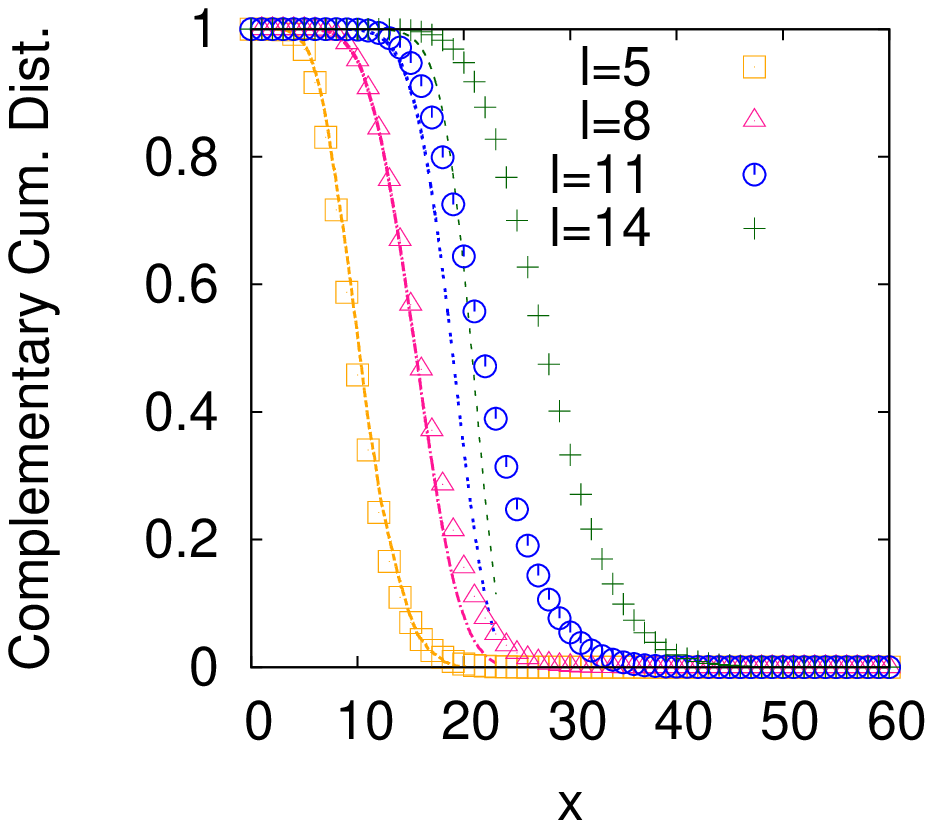}
     \begin{center} (b) $L = 10^{5}$ \end{center}
 \end{minipage} 
\caption{(Color online) CCDs of 
$x \stackrel{\rm def}{=} \log (L^{2}/A)$
restricted in the layer $l$
at time steps $t = 50, 500, 5000,$ and $50000$ from left to right, 
which correspond to $l = 5, 8, 11,$ and $14$
at the peaks of $p_{l}$ (see Fig. \ref{fig_compare_pl}) 
in this order.
The (orange, red, blue, and green) lines 
show the averaged results for 100 samples of 
the actual divisions of rectangles,
while the corresponding marks (square, triangle, circle, and plus 
from left to right)
show the results for the gamma distribution 
of Eq.(\ref{eq_gamma_dist}).} \label{fig_compare_g2l}
\end{figure}

\begin{figure}[htb]
  \begin{center}
    \includegraphics[height=73mm]{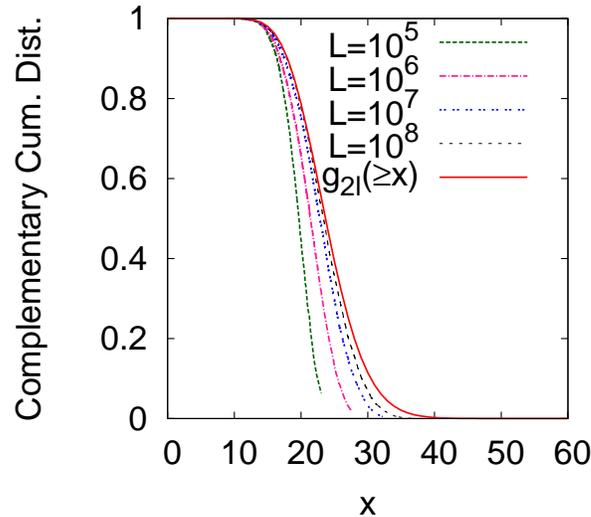}
  \end{center}
\caption{(Color online) Comparison of the CCDs of $x$
in a commonly existing layer $l = 12$ 
to the sizes of $L$ at a time step $t=500$.
From left to right, the (green, magenta, blue, and black) 
dashed lines show the averaged results for 100 samples of 
the actual divisions of rectangles
for $L = 10^{5}, 10^{6}, 10^{7},$ and $10^{8}$.
The furthest right (red) solid line shows the result for 
the gamma distribution of Eq.(\ref{eq_gamma_dist}).} 
\label{fig_gamma}
\end{figure}

Figure \ref{fig_compare_g2l} shows 
the CCD of $x = \log(L^{2}/A)$
restricted in the $l$-th layer. 
We chose the most observable layer $l = 5, 8, 11,$ and $14$, 
which correspond to the peaks of $p_{l}$
at $t= 50, 500, 5000,$ and $50000 \ll T_{max}$, respectively.
Because the most observable layer is dominant 
in the mixture distribution $\sum_{l} p_{l} g_{2l}$.
The effect of width one tends to appear as $t$ is larger 
as shown in more right curves.
The actual distributions denoted by lines and the gamma 
distribution $g_{2l}(x)$ of Eq.(\ref{eq_gamma_dist}) denoted by marks 
almost coincide 
until increasing around $t = 5000$ step (the 3rd curve from left)
in Fig. \ref{fig_compare_g2l}(a) for $L = 10^{8}$, 
however the distributions begin to differ 
from larger than 
$t = 500$ step (the 2nd curve from left) 
in Fig. \ref{fig_compare_g2l}(b) for $L = 10^{5}$.
Thus, the effect of width one becomes stronger, 
as $L$ is smaller, 
the discrepancy between the actual distribution and 
the approximative $g_{2l}$ is unignorable.
We also confirm this phenomenon for the distributions in 
a commonly existing layer 
as shown in Fig. \ref{fig_gamma}.
The small    cases of $L = 10^{5}$ and $10^{6}$ give inaccurate 
approximations even at $t=500$.
In other word, 
for a smaller $L$ at   more coarse-grained divisions,
the effect of width one already appears 
before $l=12$ in a smaller (shallower) layer.
This result is consistent with the average shrinking rate $1/e$
of edge per division, 
$e^{12} > 10^{5}$ and $e^{14} > 10^{6}$, 
as mentioned in subsection 4.2.

\begin{figure}[htb]
  \begin{center}
    \includegraphics[height=73mm]{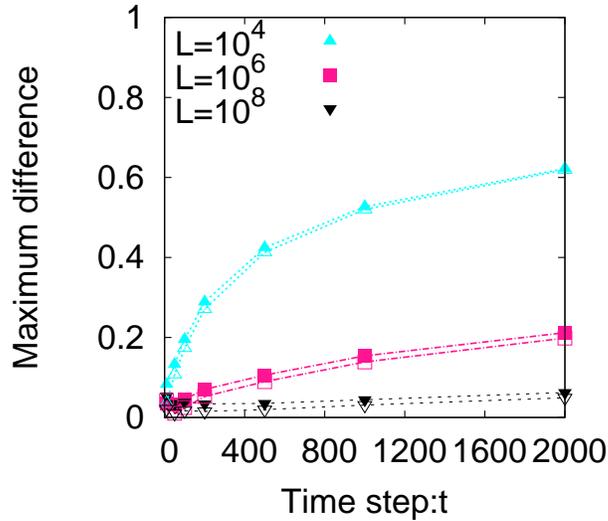}
  \end{center}
\caption{(Color online) Maximum difference in the CCDs 
$P(\geq A)$ between the averaged result by 100 samples of 
the actual divisions of rectangles
and our approximation by the mixture distribution 
$\sum_{l} p_{l} g_{2l}$ 
at several time steps $t$.
From bottom to top, the (black, magenta, and cyan)
lines with marks show the results 
for $L = 10^{8}, 10^{6}$, and $10^{4}$.
The open and closed marks correspond to the mixtures with 
the solution of Eq.(\ref{eq_difference}) 
and with the Poisson distribution of Eq.(\ref{eq_sol_pl}), 
respectively.} 
\label{fig_max-diff_cumPA}
\end{figure}

In order to grasp the tendency of approximation as a whole, 
we investigate the maximum difference in the CCD 
of areas between the actual 
data in the divisions of rectangles 
and our approximation for several time steps.
Figure \ref{fig_max-diff_cumPA} shows that 
the difference increases as $t$ is larger 
and $L$ is smaller.
In each size of $L$, 
the lines with open marks give slightly better approximation than 
the lines with closed marks,
where the former use the solution of difference equation 
(\ref{eq_difference}) and the latter use the 
Poisson distribution (\ref{eq_sol_pl}) as $p_{l}$.
The accuracy depends on the $p_{l}$ as shown in 
Fig. \ref{fig_compare_pl} at a small $t$,
while the effect of $g_{2l}$ is added 
at a large $t$ as shown in Fig. \ref{fig_compare_g2l}.
The saturated behavior in Fig. \ref{fig_max-diff_cumPA}
is not important, 
because the difference between any CCDs 
is bounded in $[0, 1]$ by nature.

\begin{figure}[htb]
 \begin{minipage}[htb]{.47\textwidth}
   \includegraphics[height=54mm]{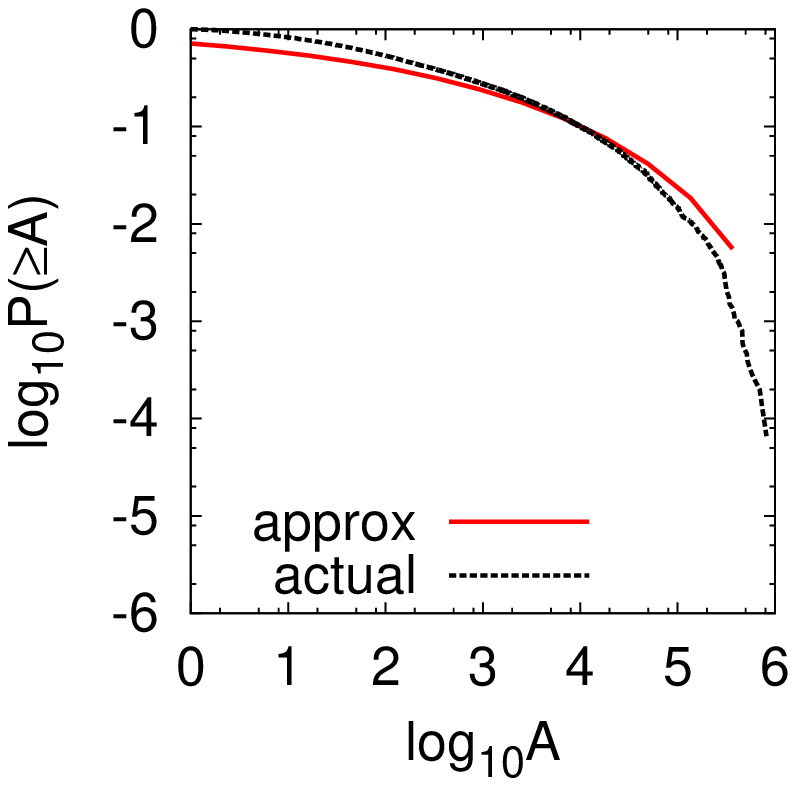}
     \begin{center} (a) Poisson, $L = 10^{3}$ \end{center}
 \end{minipage} 
 \hfill 
 \begin{minipage}[htb]{.47\textwidth}
   \includegraphics[height=54mm]{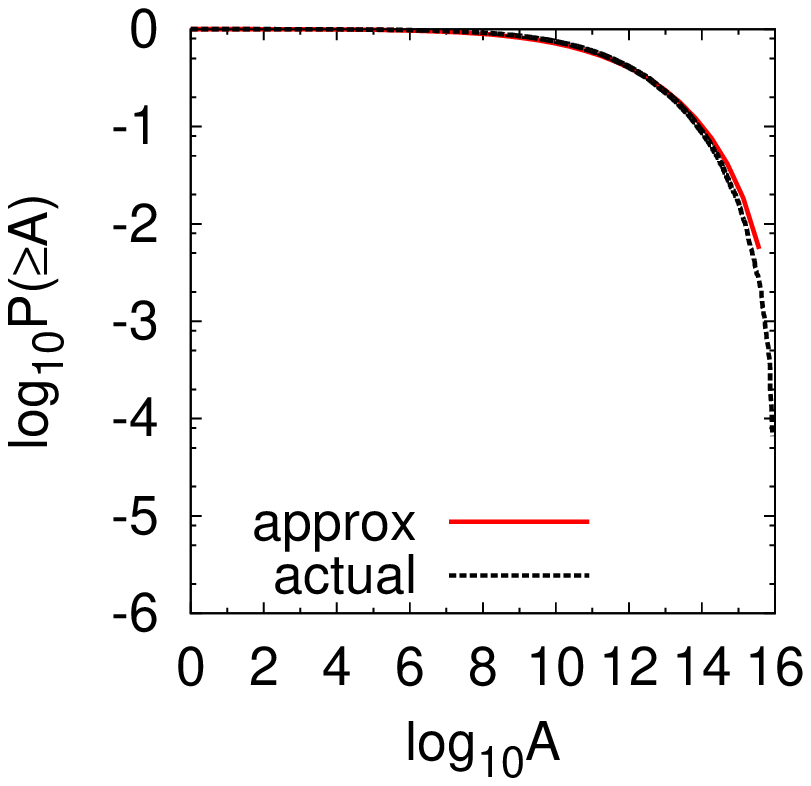}
     \begin{center} (b) Poisson, $L = 10^{8}$ \end{center}
 \end{minipage} 
 \hfill 
 \begin{minipage}[htb]{.47\textwidth}
   \includegraphics[height=54mm]{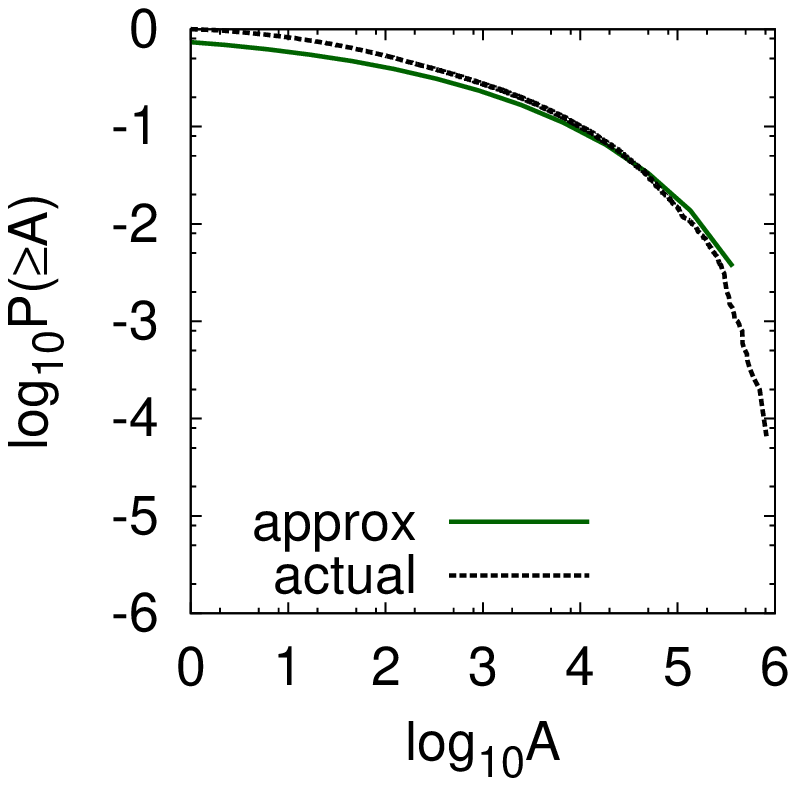}
     \begin{center} (c) Diff.Eq., $L = 10^{3}$ \end{center}
 \end{minipage} 
 \hfill 
 \begin{minipage}[htb]{.47\textwidth}
   \includegraphics[height=54mm]{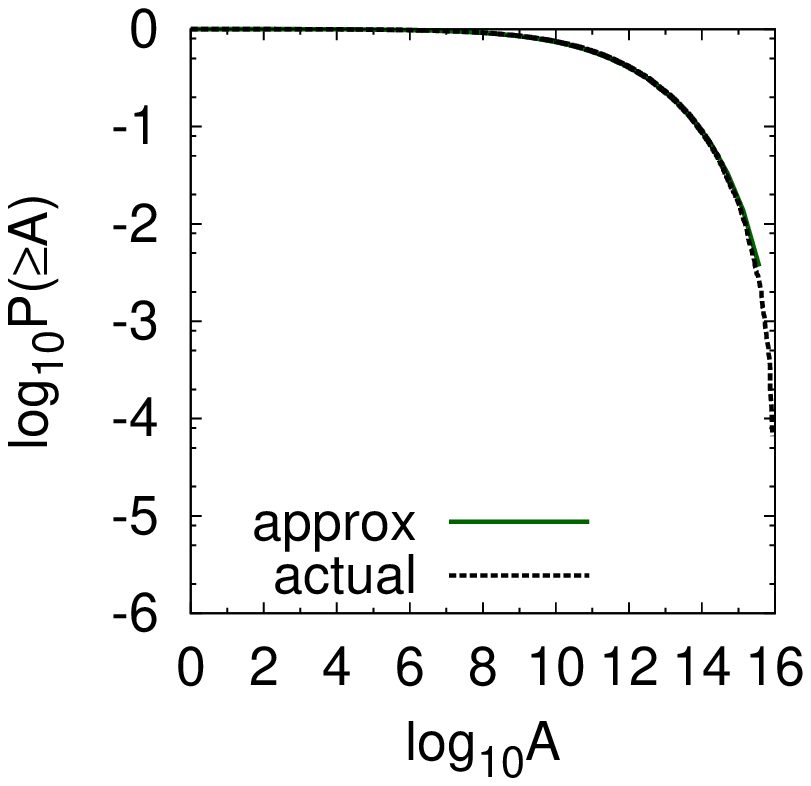}
     \begin{center} (d) Diff.Eq., $L = 10^{8}$ \end{center}
 \end{minipage} 
 \hfill 
 \begin{minipage}[htb]{.47\textwidth}
   \includegraphics[height=54mm]{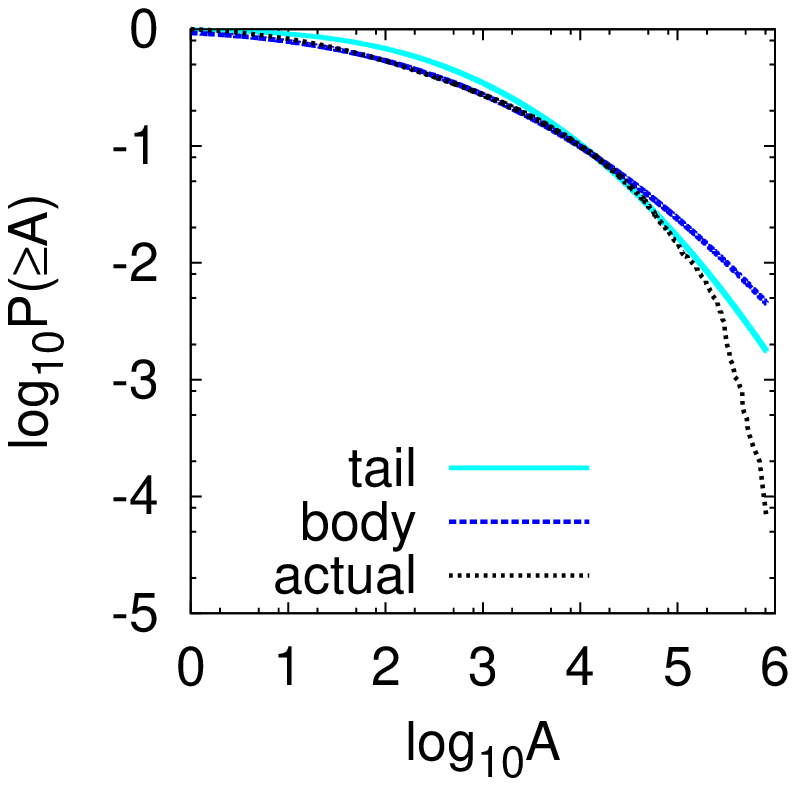}
     \begin{center} (e) lognormal, $L = 10^{3}$ \end{center}
 \end{minipage} 
 \hfill 
 \begin{minipage}[htb]{.47\textwidth}
   \includegraphics[height=54mm]{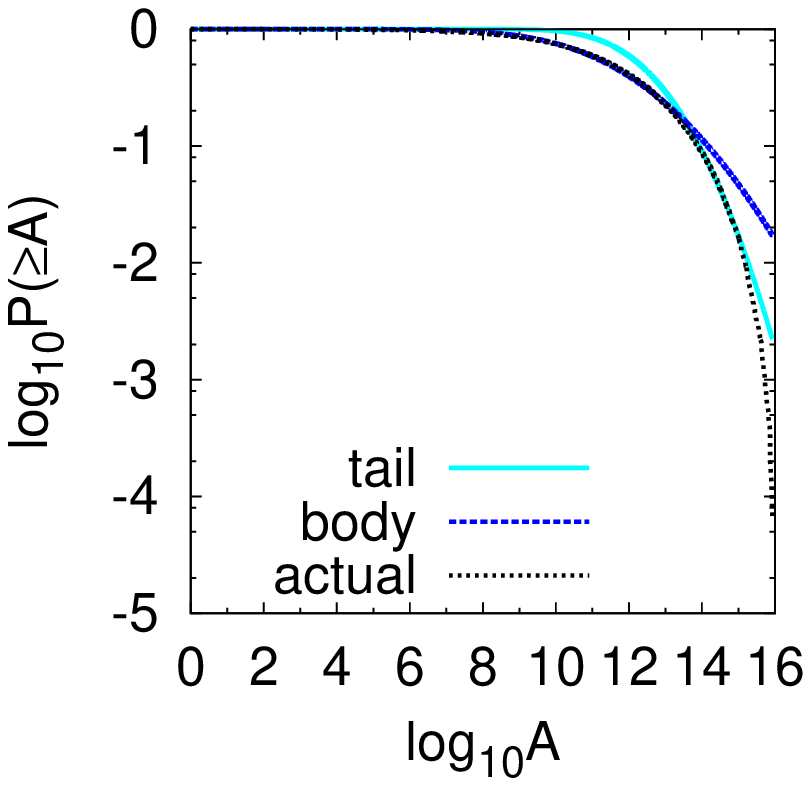}
     \begin{center} (f) lognormal, $L = 10^{8}$ \end{center}
 \end{minipage} 
\caption{(Color online) CCDs $P(\geq A)$ at $t=50$ step.
The (red or green) solid and (black) dashed lines 
show our approximation 
by using (a)(b) the Poisson distribution of Eq.(\ref{eq_sol_pl})
or (c)(d) the solution of difference equation(\ref{eq_difference})
and the averaged results by 100 samples of 
the actual divisions of rectangles, respectively.
The (blue and cyan) lines show (e)(f) CCDs 
of the estimated lognormal functions for the body 
over the whole range of $\log_{10} A$
and the tail in 
(e) $\log_{10} A \geq 5$ and (f) $\log_{10} A \geq 14$.
} \label{fig_cum_PA_T50}
\end{figure}

\begin{figure}[htb]
 \begin{minipage}[htb]{.47\textwidth}
   \includegraphics[height=54mm]{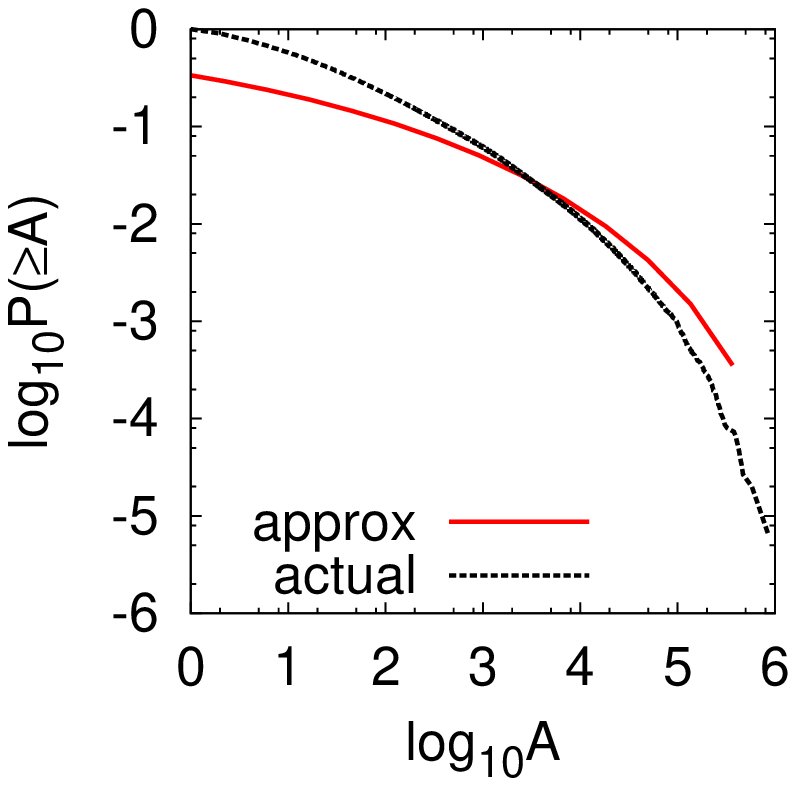}
     \begin{center} (a) Poisson, $L = 10^{3}$ \end{center}
 \end{minipage} 
 \hfill 
 \begin{minipage}[htb]{.47\textwidth}
   \includegraphics[height=54mm]{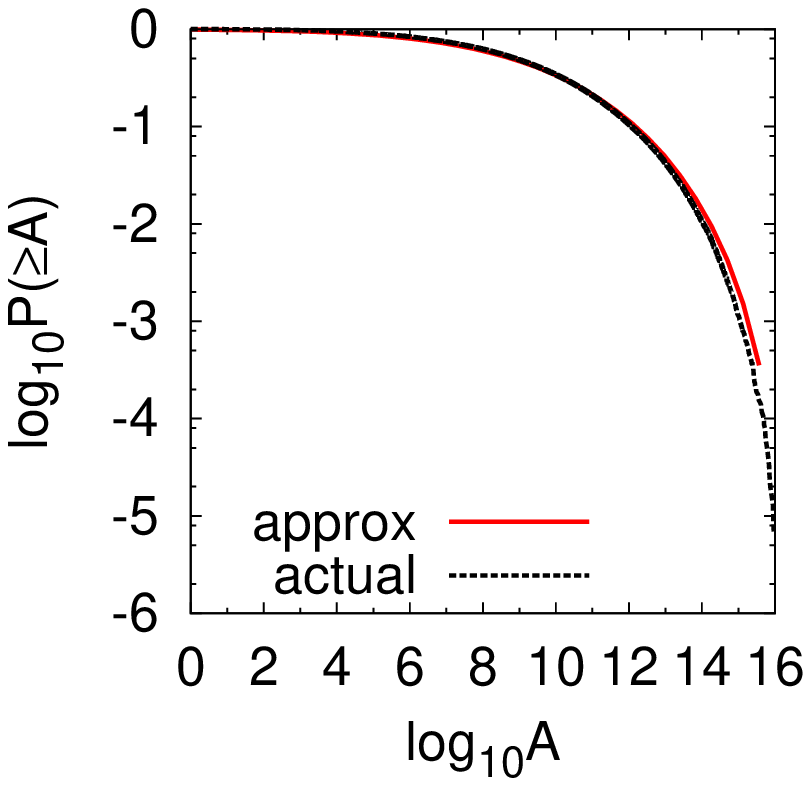}
     \begin{center} (b) Poisson, $L = 10^{8}$ \end{center}
 \end{minipage} 
 \hfill 
 \begin{minipage}[htb]{.47\textwidth}
   \includegraphics[height=54mm]{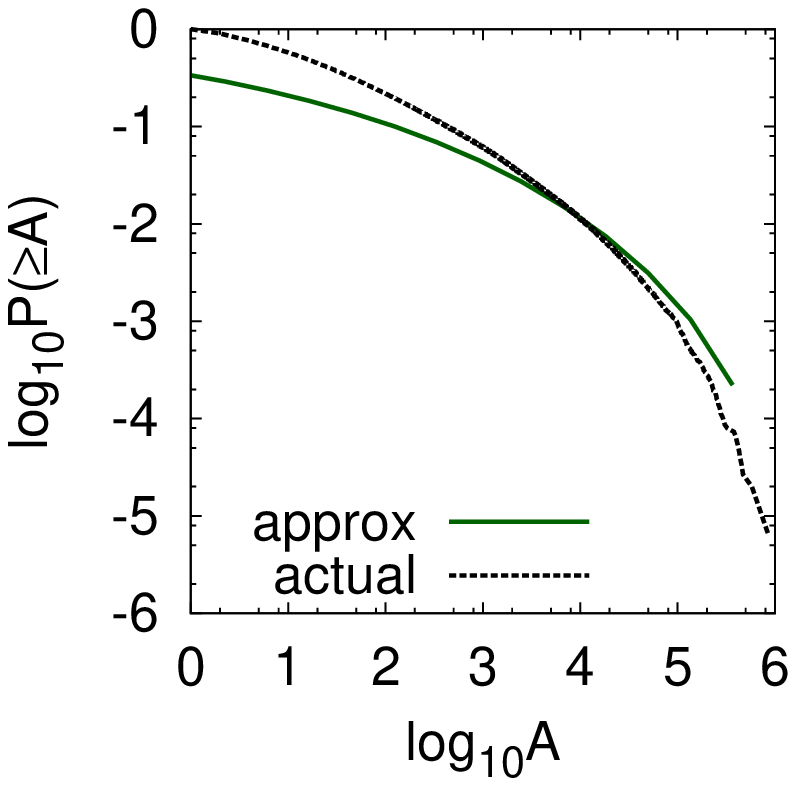}
     \begin{center} (c) Diff.Eq., $L = 10^{3}$ \end{center}
 \end{minipage} 
 \hfill 
 \begin{minipage}[htb]{.47\textwidth}
   \includegraphics[height=54mm]{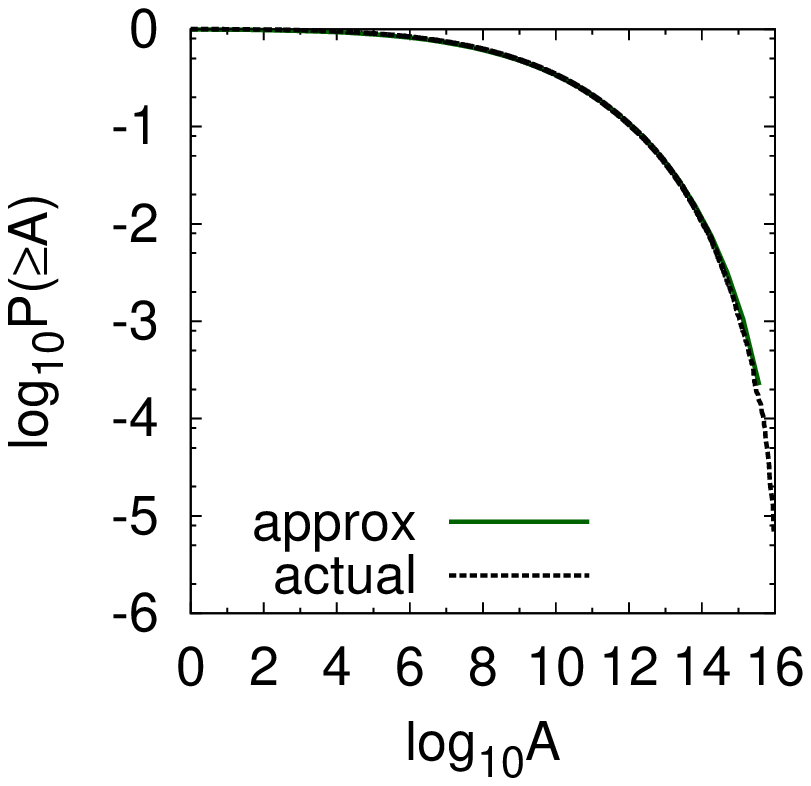}
     \begin{center} (d) Diff.Eq., $L = 10^{8}$ \end{center}
 \end{minipage} 
 \hfill 
 \begin{minipage}[htb]{.47\textwidth}
   \includegraphics[height=54mm]{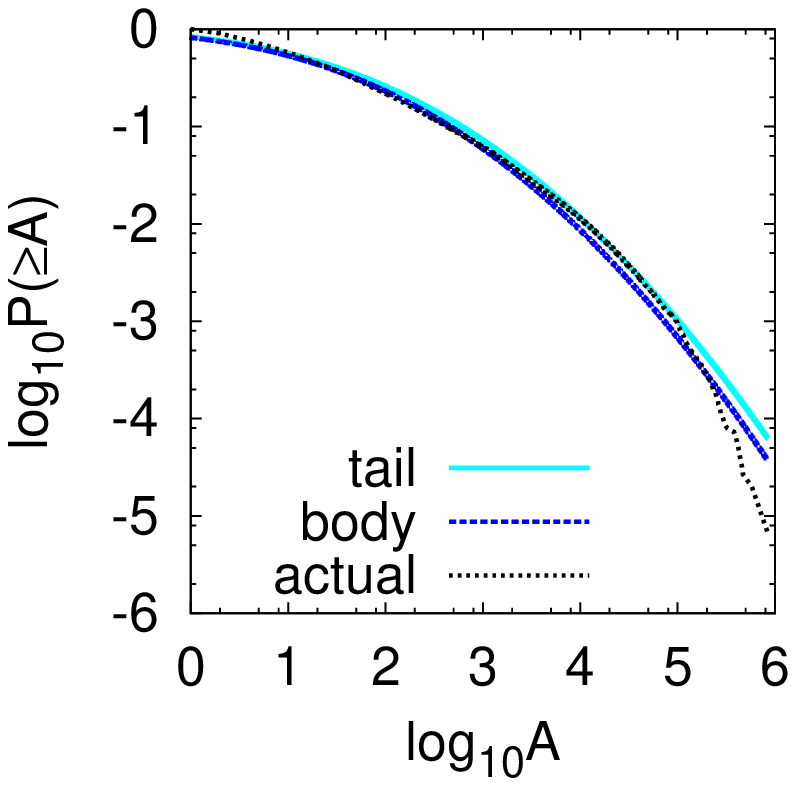}
     \begin{center} (e) lognormal, $L = 10^{3}$ \end{center}
 \end{minipage} 
 \hfill 
 \begin{minipage}[htb]{.47\textwidth}
   \includegraphics[height=54mm]{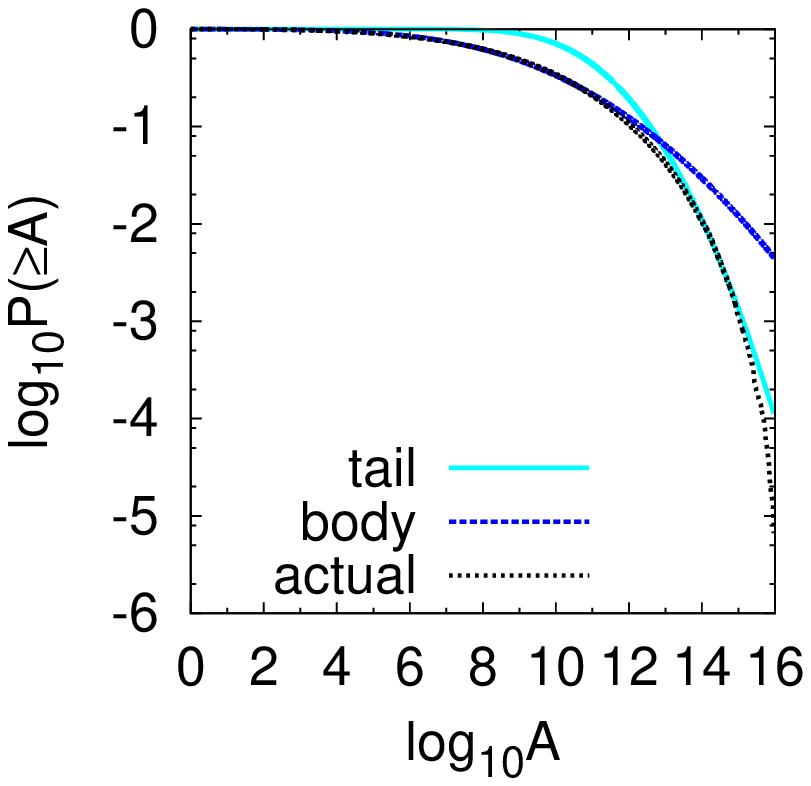}
     \begin{center} (f) lognormal, $L = 10^{8}$ \end{center}
 \end{minipage} 
\caption{(Color online) CCDs $P(\geq A)$  at $t=500$ step.
Approximation by the mixture $\sum_{l} p_{l} g_{2l}$
using (a)(b) Poisson distribution of Eq.(\ref{eq_sol_pl}) or 
(c)(d) the solution of differential Eq.(\ref{eq_difference})
as $p_{l}$, 
and (e)(f) the estimated lognormal functions 
for the body over the whole range of $\log_{10} A$
and the tail in 
(e) $\log_{10} A \geq 5$ and (f) $\log_{10} A \geq 14$.
The linetypes are the same as in Fig. \ref{fig_cum_PA_T50}.} 
\label{fig_cum_PA_T500}
\end{figure}

\begin{figure}[htb]
 \begin{minipage}[htb]{.47\textwidth} \hspace{-1cm} 
   \includegraphics[height=60mm]{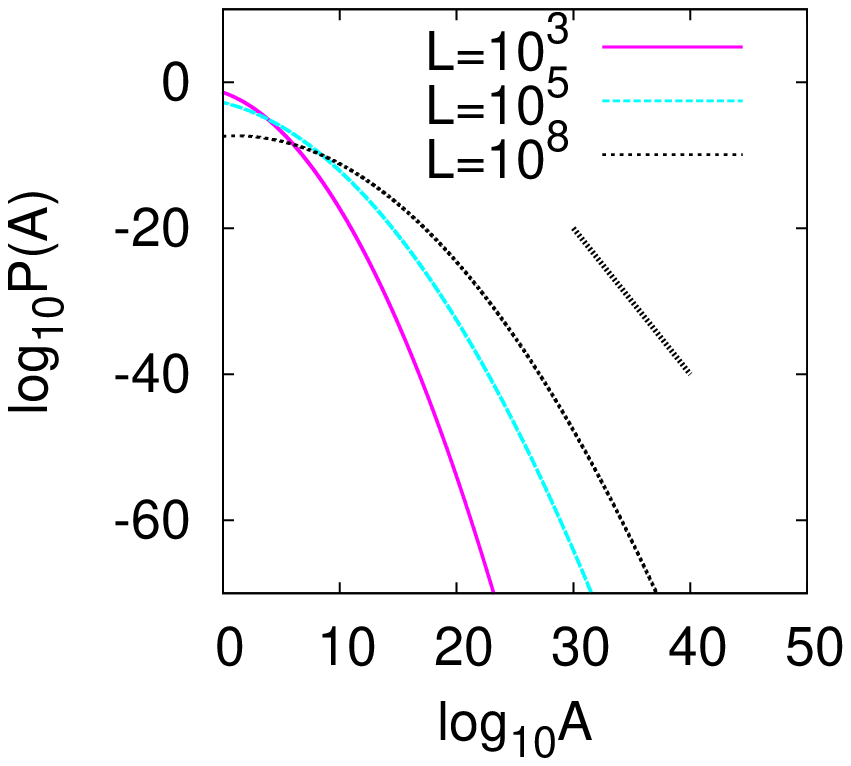}
     \begin{center} (a) Body, $t = 50$ \end{center}
 \end{minipage} 
 \hfill 
 \begin{minipage}[htb]{.47\textwidth} \hspace{-1cm} 
   \includegraphics[height=60mm]{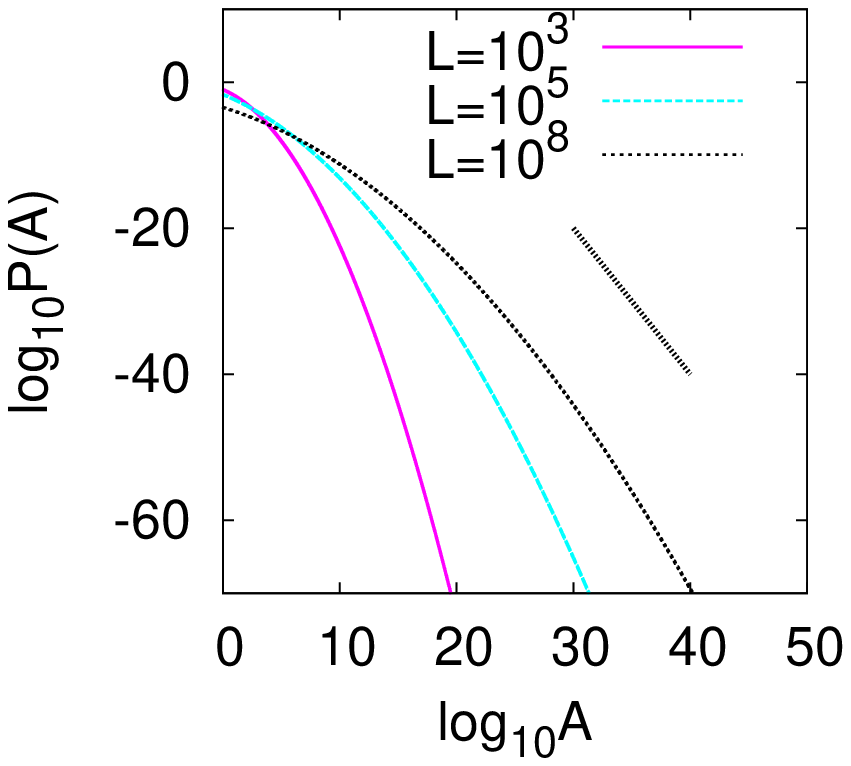}
     \begin{center} (b) Body, $t = 500$ \end{center}
 \end{minipage} 
 \begin{minipage}[htb]{.47\textwidth} \hspace{-1cm} 
   \includegraphics[height=60mm]{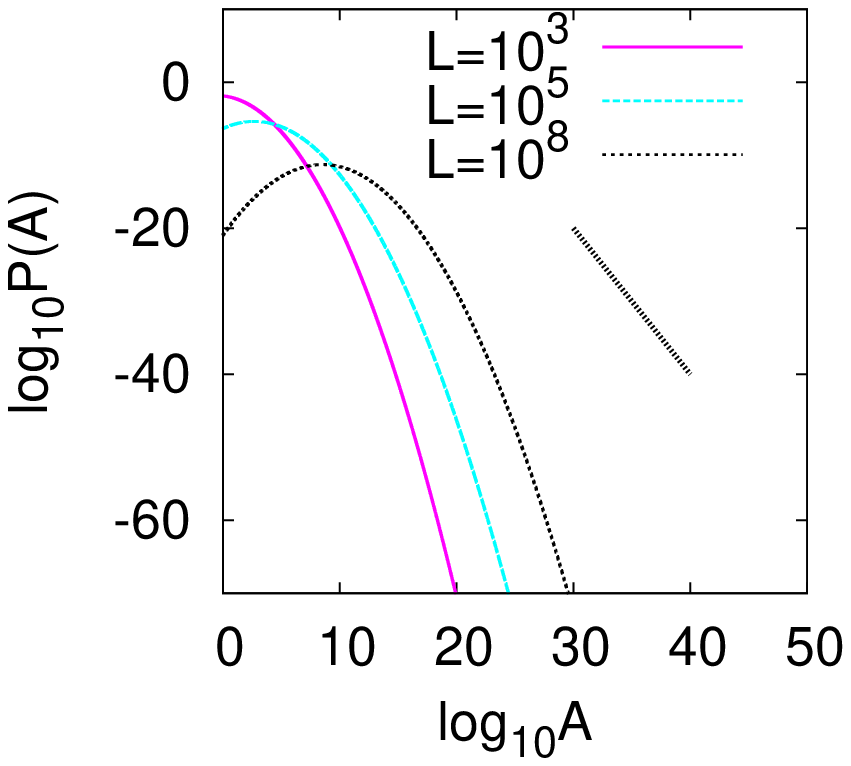}
     \begin{center} (c) Tail, $t = 50$ \end{center}
 \end{minipage} 
 \hfill 
 \begin{minipage}[htb]{.47\textwidth} \hspace{-1cm} 
   \includegraphics[height=60mm]{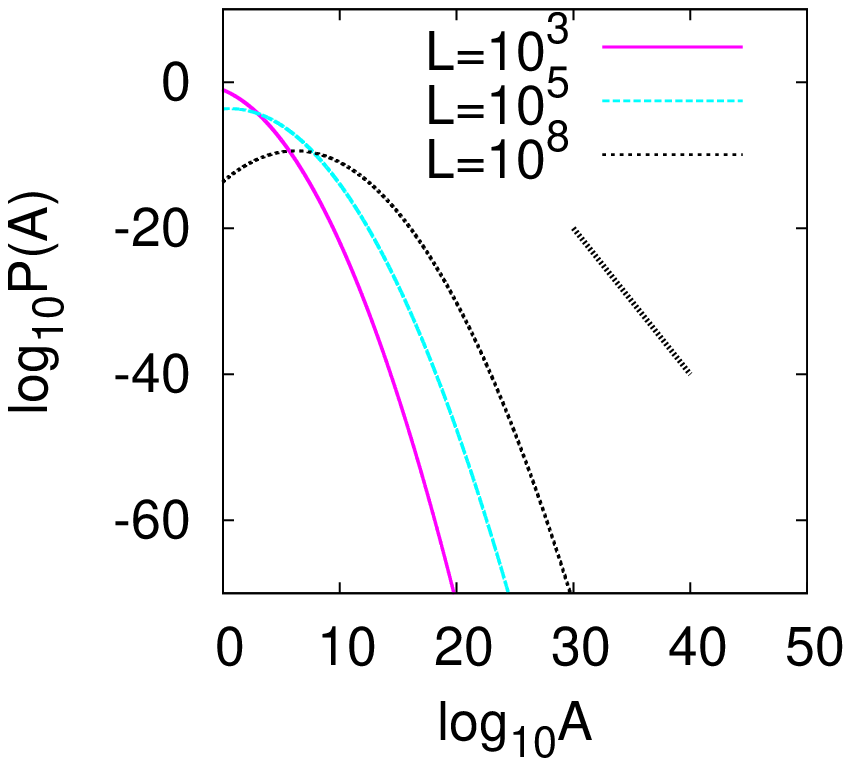}
     \begin{center} (d) Tail, $t = 500$ \end{center}
 \end{minipage} 
\caption{(Color online) Approximation of area distributions 
estimated by lognormal functions 
in Fig. \ref{fig_cum_PA_T50}(e)(f)
and Fig. \ref{fig_cum_PA_T500}(e)(f).
The short segment guides the slope of -2.} 
\label{fig_PA_T50-500}
\end{figure}

In more detail, 
we investigate the CCDs of areas 
at $t=50$ step in Fig. \ref{fig_cum_PA_T50}.
We obtain a good fitting for $L=10^{8}$ 
in Fig. \ref{fig_cum_PA_T50}(b)(d),
but remark on a small gap for $L=10^{3}$
in Fig. \ref{fig_cum_PA_T50}(a)(c).
Here, $L=10^{8}$ and $L=10^{3}$ are selected as examples 
to compare the accuracy of 
approximation affected by the appearance of 
width one in rectangles.
The discrepancy between the actual distribution 
and our approximation 
at $t=500$ step in Fig. \ref{fig_cum_PA_T500}(a)(c)
becomes larger than the corresponding results at $t=50$
in Fig. \ref{fig_cum_PA_T50}(a)(c) for $L=10^{3}$, 
while the two distributions almost coincide 
in both Figs. \ref{fig_cum_PA_T50} and  \ref{fig_cum_PA_T500}
(b)(d) for $L=10^{8}$.
Figures 
\ref{fig_cum_PA_T50}(e)(f) and \ref{fig_cum_PA_T500}(e)(f)
show reasonable fittings of the actual distribution with the
estimated lognormal functions in the cumulation
for the body over the whole range of $\log_{10} A$
and the tail restricted 
in the range of (e) $\log_{10} A \geq 5$ and 
(f) $\log_{10} A \geq 14$. 
For the body and the tail, 
Figure \ref{fig_PA_T50-500} shows the 
lognormal density functions.
The short segment represents the slope of 
the exponent $-2$ 
for fragment-mass of glass \cite{Katsuragi04}
and areas enclosed by the city roads \cite{Lammer06}\cite{Masucci09}.
The linear part in Fig. \ref{fig_PA_T50-500} 
seems to be longer for larger $L$ further to the right, 
as similar to the area distributions of extreme rectangles 
in Fig. \ref{fig_limit_dist}.
Remember that a large $L$ gives fine-grained divisions.
This phenomenon may be consistent with 
the enhancement of power-law property 
as high impact in the fragmentation of glass \cite{Katsuragi04},
however the behavior in our model is not exactly the same 
because of the remaining lognormal property 
whose distribution rather resembles to the ones for road networks 
in German cities \cite{Chan11}.
We note that  
the approximation is no longer accurate for $L=10^{8}$
in $t \gg 500$. 
The maximum difference $< 0.1$ in Fig. \ref{fig_max-diff_cumPA}
may be a criterion for whether it gives a good fitting or not.

\newpage
\section{Conclusion} \label{sec5}
Fat-tail distributions are pervasive in nature,
and also appear  in spatial networks. 
In particular, lognormal and power-law distributions are familiar 
in fragments of glass \cite{Katsuragi04,Ishii92}, 
crack patterns, and 
areas enclosed by city roads \cite{Lammer06,Masucci09,Chan11}.
Beyond the details of physical phenomena, 
it is useful to consider a common generation mechanism 
on the division process.
Thus, 
we have investigated a simple model
for producing such distributions of areas enclosed by edges 
in a spatial network,
which is 
an extension of MSQ networks \cite{Hayashi09,Hayashi10,Hayashi11}
from the iterative divisions of squares to ones of rectangles.

The stochastic division process makes a Markov chain
in the random selections of a rectangle face and of the 
division axes from the initial configuration of an 
$L \times L$ square.
We have derived the exact solution of distribution at $T_{max}$
for the extreme rectangles with no more divisible edge(s) 
of width one on a combinatorial analysis for a small $L$.
It is also pointed out that, in the absorbing Markov chain,  
the iterative divisions of rectangles are equivalent 
to directional random walks with splitting into four particles.
For a large $L$ and $t \ll T_{max}$, 
we have discussed the distribution of areas
on a continuous approximation.
As the distributions of layers and of areas restricted in a layer, 
we decompose the original distribution into two functions 
and consider the mixture of them, 
which corresponds to the Poisson and the gamma distributions
in Eqs. (\ref{eq_sol_pl}) and (\ref{eq_gamma_dist}).
In addition, 
we obtain the distribution of layers 
by the difference equation (\ref{eq_difference})
more accurately.
Simulation results show a good agreement of our approximation with 
the actual distribution in the divisions, 
and give a condition for the fitting.
We obtain more accurate results, 
as the size $L$ is larger and the time step $t$ is smaller,
since 
the layer of a face tends to be shallower and 
the effect of width one in rectangles becomes weaker.
We emphasize that 
the decomposition into two distributions of layers and of 
areas restricted in a layer will be useful for investigating 
other phenomena, such as in broken 
fragments of glass and areas enclosed by city roads,
with additional information of layers defined by the 
time sequence of divisions.

We also confirm a slightly better agreement of our approximation 
for the $m=2$ divisions which correspond to a crack model, 
though the meaning of network construction by bridgings 
should be discussed further from an application point of view.
Unfortunately, 
relations to the mathematical properties of STIT tessellations 
\cite{Nagel05}\cite{Nagel09}\cite{Thale09}
are unknown.
In addition, there remain further challenges 
to the analysis for other properties, 
e.g., the lifetime of face 
or the distribution of edge lengths 
in our framework of stochastic processes, 
more rigorous investigation for the fitting functions
(e.g., estimation by a double Pareto 
\cite{Mitzenmacher04a}\cite{Reed04}), 
extension to a preference selection model, and 
considering the division by 
any direction not limited to vertical and horizontal.

\section*{Acknowledgment}
The authors would like to thank Prof. Mitsugu Matsushita 
(Chuo University) for suggesting a similarity 
with the property for fragmentation of glass 
in Refs. \cite{Katsuragi04} and \cite{Ishii92}. 
This research is supported in part by a 
Grant-in-Aid  for Scientific Research in Japan, No. 21500072.

\section*{Appendix A: Fractal dimension}
We consider the fractal dimension $d_{f}$ 
of a MSQ network \cite{Hayashi10} 
based on the self-similar tiling by squares at a finite 
time $\tau < \infty$. 
Note that 
the asymptotical value is $d_{f} \rightarrow 2$ 
trivially for the infinite time $\tau \rightarrow \infty$, 
since the division process has no limitation and 
the whole two-dimensional space is embedded by the 
recursive subdivisions after a very long time.

In general, for the number $N_{b}[l]$ of covered boxes 
by a measure $\epsilon = 2^{-l}$, 
it is defined as 
\[
  d_{f} = \lim_{l \rightarrow \infty} 
	\frac{\log N_{b}[l]}{\log (1/\epsilon)}
	= \lim_{l \rightarrow \infty} 
	\frac{\log N_{b}[l]}{l \cdot \log 2}.
\]

For each visible level, the number $N_{b}[l]$ 
is counted as 
\[
\begin{array}{ccc}
  N_{b}[2] & = & 2 \times N_{b}[1] + n_{2}(\tau),\\
  N_{b}[3] & = & 2 \times N_{b}[2] + n_{3}(\tau),\\
    :         & & 
\end{array}
\]
where, in the above right-hand sides, 
the first term is due to the doubling of measure, 
and the second term comes from the one-to-one 
correspondence between four faces and cross edges 
generated by the quadratic division 
(e.g., in a clockwise mapping). 
The recursion is 
$N_{b}[l] = 2^{l-1} N_{b}[1] 
+ \sum_{i = 2}^{l} 2^{l-i} n_{i}(\tau)$ 
with $N_{b}[1] = 12$ 
in an initial configuration of four squares.
By substituting Eq.(\ref{eq_sol_nl}) with $m=4$ 
into $\sum_{i = 2}^{l} 2^{l-i} n_{i}(\tau)$,
we derive 
\[
 \begin{array}{lcl}
   \sum_{i = 2}^{l} 2^{l-i} 4^{i-1} 
	\frac{\tau^{i-1}}{(i-1)!} e^{- \tau}
	& = & 2^{l} \sum_{i = 2}^{l} 
	\frac{(2 \tau)^{i-1}}{2 (i-1)!} e^{- \tau},\\
	& = & 2^{l} (e^{2 \tau} -1 -Res) e^{-\tau}/2,
 \end{array}
\]
where $Res$ denotes the residual for the higher-terms than 
$l$ in the Taylor expansion of $e^{x}$.

For $l \gg 1$, we obtain 
\begin{equation}
  \log N_{b}[l] \sim \log (2^{l} (6 + e^{\tau}/2) ) 
	\sim l \cdot \log 2 + \tau.
\label{eq_logN}
\end{equation}

On the other hand, 
by using a generating function
$F_{t}(z) \stackrel{\rm def}{=} 
\sum_{w \in Leaves_{t}} z^{\mid w \mid}$, it is also represented as 
\[
  N_{b}[l] = 4 \times \
	\sum_{w \in Leaves_{t}} 2^{l - \mid w \mid}
	= 4 \times 2^{l} F_{t}(1/2), 
\]
where $\mid w \mid$ denotes the depth of face $w$, and 
$Leaves_{t}$ is a set of faces at the division of $t$-step 
in the random quadtree, which corresponds 
to our hierarchical network model.
From the LEMMA 7.1 in \cite{Eisenstat11}, 
the expectation is 
\[
  {\cal E}(N_{b}[l]) = 4 \times 2^{l} 
	{\cal E}(F_{t}(1/2)),
\]
\[
  {\cal E}(F_{t}(z)) \leq \exp \left( 
	(4 z -1) \sum_{k=0}^{t-1} \frac{1}{3k + 1} \right).
\]
When we set $3k+1 = x$, 
\[
  {\cal E}(F_{t}(1/2)) \leq \exp \left(
	\int_{1}^{3t-2} \frac{1}{3x} dx \right)
	\leq (3t+1)^{1/3}.
\]

From the relation $1+3t = e^{3 \tau}$ of the total number of faces, 
we obtain 
\[
  \log {\cal E}(N_{b}[l]) \leq \log 4 + l \cdot \log 2 + 
	\frac{1}{3} \log(1+3t) \sim l \cdot \log 2 + \tau.
\]
This is equivalent to Eq.(\ref{eq_logN}).

We consider a typical time 
$\langle \tau \rangle = \sum_{\tau} \tau p_{l}(\tau) \times m$
for the layer $l$,
\[
  \langle \tau \rangle \approx m
	\int_{0}^{\infty} \frac{m^{l-1}}{(l-1)!} 
	\tau^{l} e^{-m \tau} d \tau 
	= \frac{l}{m}, 
\]
where the factor $m$ is due to the normalization, 
\[
  \sum_{\tau} p_{l}(\tau) \approx 
	\frac{1}{(l-1)!} \int_{0}^{\infty} (m \tau)^{l-1} 
	e^{-m \tau} d \tau
	= \frac{1}{m}.
\]
Thus, as an upper-bound for a large $l$, 
we obtain 
\[
  d_{f} \sim 
	\frac{l \cdot \log 2 + l/m}{l \cdot \log 2}
	= 1 + \frac{1}{m \cdot \log 2} = 1.36067.
\]
Note that 
this value is between Koch curve: $\log 4/\log 3 = 1.26186$ 
and 
Sierpinski carpet: $\log 8/ \log 3 = 1.89278$ or 
Sierpinski gasket: $\log 3/ \log 2 = 1.58496$.

\section*{Appendix B: Extended preference model}
When a rectangle face is chosen (not uniformly at random but) 
proportionally to the power $\gamma^{l}$ of its depth $l$
with a real parameter $\gamma > 0$, 
we can approximately derive the distribution 
$p_{l} = n_{l} / {\cal N}$.
Note that, in the selected division, 
a larger face (at a shallower layer) is preferred 
for $\gamma < 1$, 
while a smaller face (at a deeper layer) is preferred 
for $\gamma > 1$  \cite{Eisenstat11}.

The number $n_{l}$ of $l$-th faces at time $\tau$ follows 
\begin{eqnarray}
  \frac{d n_{l}}{d \tau} & = & 
	m \frac{\gamma^{l-1}}{C} n_{l-1} 
	- \frac{\gamma^{l}}{C} n_{l}, \;\;\; l \geq 2, 
	\label{eq_r-depend}
\\
  \frac{d n_{1}}{d \tau} & = & - \frac{r^{1}}{C} n_{1}.
	\label{eq_r0}
\end{eqnarray}

As $\gamma \approx 1$,
we assume that $1/C$ is a constant.
The reason is discussed later. 
By a variable transformation $\tau = C \gamma^{-l} T$, 
we rewrite Eq.(\ref{eq_r-depend}) as
\[
  \frac{\gamma^{l}}{C} \frac{d n_{l}}{d T} 
	= \frac{d n_{l}}{d \tau} 
	= \frac{\gamma^{l}}{C} 
	\left( \frac{m}{\gamma} n_{l-1} - n_{l} \right).
\]
Thus, 
we obtain the same form of Eq.(\ref{eq_diff_tau}) 
at $\gamma=1$ as follows
\[
  \frac{d n_{l}}{d T} = \frac{m}{\gamma} n_{l-1} - n_{l}.
\]
The solution is 
\[
  n_{l} = \left( \frac{m}{\gamma} \right)^{l-1} 
	\frac{T^{l-1}}{(l-1)!} e^{-T}
	= \left( \frac{m}{\gamma} \right)^{l-1} 
	\frac{(\gamma^{l} \tau/C)^{l-1}}{(l-1)!} e^{-\gamma^{l}\tau/C}, 
\]
\[
  {\cal N} = \sum_{l=1} n_{l} = 
	\sum_{l=1} \frac{(m \gamma^{l-1} \tau/C)^{l-1}}{(l-1)!} 
	e^{- \gamma^{l} \tau/C}.
\]

Here, we evaluate the assumption of a constant $1/C$. 
Since the number of faces increases by $m-1 = 3$
(add four divisions, and delete one chosen face)
at each time step $t$, 
the total number of faces is 
$\sum n_{l} = 1+(m-1)t$. 
It must be equal to Eq(\ref{eq_sol_N}),
so that $\sum n_{l} = e^{(m-1)\tau}$. 
By applying 
$d \sum n_{l} / d \tau = (m-1) \sum n_{l}$, 
the following relation must hold 
from Eqs.(\ref{eq_r-depend}) and (\ref{eq_r0}),
\[
 \sum_{l=1} \frac{d n_{l}}{d \tau} 
	= \frac{m-1}{C} \sum_{l=1} \gamma^{l} n_{l} 
	= (m-1) e^{(m-1)\tau}, 
\]
therefore 
$C = (\sum_{l} \gamma^{l} n_{l}) e^{-(m-1)\tau}$. 
Indeed, around $\gamma \approx 1$, 
$1/C$ is almost constant as 
approximated in Table \ref{table_invC}. 

\begin{table}[htb]
\begin{center}
\begin{tabular}{c|cccc} \hline
$\tau$   & 1    & 2     & 3     & 4 \\ \hline
$\gamma=0.95$ & 1.21 & 1.42 & 1.66 & 1.93 \\
$\gamma=0.97$ & 1.12 & 1.25 & 1.38 & 1.52 \\
$\gamma=0.99$ & 1.04 & 1.08 & 1.12 & 1.16 \\ \hline
$\gamma=1.01$ & 0.96 & 0.92 & 0.88 & 0.84\\
$\gamma=1.03$ & 0.88 & 0.75 & 0.61 & 0.52\\
$\gamma=1.05$ & 0.77 & 0.46 & 0.34 & 0.29\\ \hline
\end{tabular}
\end{center}
\caption{The estimated values of $1/C$ by the Newton-Raphson method
for varying a parameter $\gamma$. 
Note that $\tau$ takes a small value even for a huge network 
because $\tau$ is a logarithmic timescale for 
a linear timescale $t = 1, 2, \ldots$ of discrete steps 
in the relation $1+(m-1)t = e^{(m-1)\tau}$.}
\label{table_invC}
\end{table}



\begin{thebibliography}{100}
\vspace{-4mm}
\bibitem{Simon55}
H.A. Simon, 
{\em Biometrika}, 
{\bf 42}, 425--440, (1955).


\bibitem{Newman06} M.E.J. Newman, A.-L. Barab\'{a}si, and
D.J. Watts,
Chapter Three: Empirical Studies, pp.167--228, 
The Structure and Dynamics of NETWORKS,
Princeton University Press, 2006.

\bibitem{Mitzenmacher04a}
M. Mitzenmacher, 
{\em Internet Mathematics}, 
{\bf 1(2)}, 226--251, (2004).
\bibitem{Mitzenmacher04b}
M. Mitzenmacher, 
{\em Internet Mathematics}, 
{\bf 1(3)}, 305--333, (2004).
\bibitem{Reed04}
W.J. Reed, and M. Jorgensen, 
{\em Communications in Statistics: Theory and Methods}, 
{\bf 33(8)}, 1733-1753, (2004).

\bibitem{Katsuragi04}
H. Katsuragi, D. Sugino, and H. Honjo, 
{\em Phys. Rev. E 70}, 065130, (2004).
\bibitem{Ishii92}
T. Ishii, and M. Matsushita, 
{\em J. of The Physical Society of Japan}, 
{\bf 61(10)}, 3474--3477, (1992).

\bibitem{Lammer06} S. L\"{a}mmer, B. Gehlsen, D. Helbing, 
{\em Physica A} {\bf 363}, 89, (2006).
\bibitem{Masucci09} A.P. Masucci, D. Smith, and C.M. Batty, 
{\em Eur. Phys. J. B} {\bf 71(2)}, 259, (2009).
\bibitem{Chan11} 
S.H.Y. Chan, R.V. Donner, and S. L\"{a}mmer, 
{\em Eur. Phys. J. B} {\bf 84(4)}, 563--577, (2011).

\bibitem{Delaney08} 
G.W. Delaney, S. Hutzler, and T. Aste,
{\em Phys. Rev. Lett.} {\bf 101}, 120602, (2008).
\bibitem{Dodds03} 
P.S. Dodds, and J.S. Weitz, 
{\em Phys. Rev. E} {\bf 67}, 016117, (2003).


\bibitem{Barthelemy08} 
M. Barthelemy, and A. Flammini,
{\em Phys. Rev. Lett.} {\bf 100}, 138702, (2008).

\bibitem{Lee12} 
S.-H. Lee, and P. Holme,
arXiv:1205.0537, (2012).


\bibitem{Nagel05}
W. Nagel, and V. Weiss, 
{\em Adv. Appl. Prob.(SGSA)}, 
{\bf 37}, 859--883, (2005).

\bibitem{Nagel09}
W. Nagel, J. Mecke, J. Ohser, and V. Weiss, 
{\em Image Anal. Stereol.}, 
{\bf 27}, 73--78, (2008).

\bibitem{Thale09}
C. Th\"{a}le, 
{\em Image Anal. Stereol.}, 
{\bf 28}, 69--76, (2009).


\bibitem{Kalapala06}
V. Kalapala, V. Sanwalani, A. Clauset, and C. Moore, 
{\em Phys. Rev. E} {\bf 73}, 026130, (2006).

\bibitem{Cowan10} R. Cowan, 
{\em Advances in Applied Probability} {\bf 42(1)}, 26--47, (2010).
\bibitem{Eisenstat11} D. Eisenstat,
{\em Proceeding of SIAM the 8th Workshop on 
Analytic Algorithms and Combinatorics} (ANALCO11), Jan. 22, 2011. 

http://arxiv.org/abs/1008.4916
http://www.siam.org/proceedings/analco/2011/anl11\_09\_eisenstatd.pdf

\bibitem{Hayashi09} Y. Hayashi, 
{\em Physica A} {\bf 388}, 991, (2009).
\bibitem{Hayashi10}
Y. Hayashi, and Y. Ono, 
{\em Phys. Rev. E 82}, 016108, (2010).
\bibitem{Hayashi11} Y. Hayashi, 
{\em IEICE Trans. Fundamentals}, {\bf E94-A(2)}, 846-849, (2011).

\bibitem{Karavelas01}
M.I. Karavelas, and L.J. Guibas,
{\em Proc. of the 12th ACM-SIAM Symposium
on Discrete Algorithms}, (2001).
















\bibitem{Liggett99} T.M. Liggett,
Stochastic Interacting Systems: Contact, Voter and Exclusion Processes,
Springer, 1999.
\bibitem{Derrida93} B. Derrida, M.R. Evans, V. Hakim, and V. Pasquier, 
{\em J. Phys. A} {\bf 26}, 1493, (1993).




\end{thebibliography}
\end{document}